\documentclass[12pt]{article}
\usepackage{amsmath}
\usepackage{graphicx}
\usepackage{enumerate}
\usepackage{natbib}
\usepackage{url} 

\newcommand{\blind}{1}

\RequirePackage[OT1]{fontenc}
\RequirePackage{amsthm,amsmath}
\RequirePackage[colorlinks,citecolor=black,urlcolor=blue]{hyperref}
\hypersetup{backref, final=false, pdfpagemode=FullScreen,  colorlinks=true}
\usepackage{graphicx}
\usepackage{amssymb}
\usepackage[ruled]{algorithm2e}
\usepackage{color,soul}
\usepackage{multirow}
\usepackage{amsthm,amsmath,amssymb}
\usepackage{mathrsfs}
\usepackage{color,hyperref}
 \makeatletter
 \def\singlespace{\def\baselinestretch{0.7}\@normalsize}
\def\beqn{\begin{eqnarray}}
\def\eeqn{\end{eqnarray}}
\def\beqns{\begin{eqnarray*}}
\def\eeqns{\end{eqnarray*}}

\newtheorem{theorem}{Theorem}

\newtheorem{example}{Example}

\renewcommand{\theequation}{\arabic{equation}}
\newcounter{thm}

\newcounter{lem}

\def\bmu{\mbox{\boldmath{$\mu$}}}

\def\bomega{\mbox{\boldmath{$\omega$}}}

\def\bSigma{\mbox{\boldmath{$\Sigma$}}}

\def\bGamma{\mbox{\boldmath{$\Gamma$}}}
\def\bdelta{\mbox{\boldmath{$\delta$}}}

\def\bX{\mathbf X}

\def\bD{\mathbf D}
\def\mD{\mathcal D}
\def\bE{\mathbf E}
\def\be{\mathbf e}
\def\mF{\mathbf F}
\def\bG{\mathbf G}

\def\mH{\mathcal H}

\def\bL{\mathbf L}
\def\bP{\mathbf P}

\def\bY{\mathbf Y}
\def\bb{\mathbf b}
\def\bc{\mathbf c}

\def\bv{\mathbf v}

\def\b0{\mathbf 0}

\def\bPi{\mbox{\boldmath{$\Pi$}}}

\def\bmu{\mbox{\boldmath{$\mu$}}}

\def\bomega{\mbox{\boldmath{$\omega$}}}

\def\bSigma{\mbox{\boldmath{$\Sigma$}}}

\def\bGamma{\mbox{\boldmath{$\Gamma$}}}
\def\bdelta{\mbox{\boldmath{$\delta$}}}

\def\Var{\mbox{\rm Var}}

\begin{document}

\def\spacingset#1{\renewcommand{\baselinestretch}%
{#1}\small\normalsize} \spacingset{1}


\if1\blind
{
  \title{\bf Nonnested model selection based on empirical likelihood}
  \author{Jiancheng Jiang\\
    Department of Mathematics and Statistics
\& School of Data Science, \\ University of North Carolina at Charlotte, NC 28223, USA.\\\\
    Xuejun Jiang \thanks{
   Correspondence: Xuejun Jiang, Department of Statistics and Data Science, Southern University of Science and Technology, Shenzhen, 518055, China. Email: \it{jiangxj@sustech.edu.cn}}\hspace{.2cm}\\
    Department of Statistics and Data Science,\\
Southern University of Science and Technology, Shenzhen, 518055, China.\\\\
     Haofeng Wang \\
    Department  of Mathematics,\\
Harbin Institute of Technology, Harbin, 150001, China.}
  \maketitle
} \fi

\if0\blind
{
  \bigskip
  \bigskip
  \bigskip
  \begin{center}
    {\LARGE\bf Nonnested model selection based on \\ [0.5cm]
    empirical likelihood}
\end{center}
  \medskip
} \fi

\bigskip
\begin{abstract}
We propose an empirical likelihood ratio test for nonparametric model selection,
where the competing models  may be nested, nonnested, overlapping, misspecified, or correctly specified.
It compares the squared prediction errors of models based on the cross-validation and
allows for heteroscedasticity of the errors of models.
We develop its asymptotic distributions for comparing additive models and varying-coefficient models
and extend it to test significance of variables in additive models with massive data.
The method is applicable to other model comparison problems.
To facilitate implementation of the test, we provide a fast calculation procedure.
Simulations show that the proposed tests work well and have favorable finite sample performance over some existing approaches.
The methodology is validated on an empirical application.
\end{abstract}

\noindent%
\vfill

\newpage
\spacingset{1.9} 
\addtolength{\textheight}{.5in}%
\section{Introduction}
\label{sec1}

In application, one often needs to decide which model works better for a given dataset among a set of misspecified models since
no model is right. This motivates us to introduce a novel empirical likelihood ratio (\textsc{elr})
test to model selection.
The proposed method is applicable to model selection between any two supervising learning models, which may be nested, nonnested, overlapping, misspecified, or correctly specified.

Most existing model selection methods use likelihood or information criteria, such as \textsc{aic}, \textsc{bic}, \textsc{lasso} or \textsc{scad}, etc. They are widely used in statistical theory and have made great success in practice, but cannot be directly applied to nonnested model selection. Consider, for example, selecting important genes in the non-Hodgkin's lymphoma data in Dave et al. (2004) using the famous Cox's model and the additive hazard model, based on the \textsc{lasso}. Each model may lead to a different group of important genes, but there is no general tool to judge which model is better. Since the two models are nonnested, the likelihood comparison does not make sense, and hence the \textsc{aic} and \textsc{bic} criteria cannot be used.
Comparison of nonnested models also arises in time series modeling, for instance, assessment of an ARCH(7) model versus a GARCH(1,2) model.
In other situations, even if models are nested, one may have difficulty in making a decision on selecting a better model. For example, suppose there are two candidate models with \textsc{aic} values equal to 100 and 102. Then the model with an \textsc{aic} value of 100 is preferred according to this criterion. However, one cannot conclude that it is definitely better,
 because one \textsc{aic} value is smaller than the other possibly due to randomness of the sample.
 In other words, one does not have a clear cutoff for the difference of \textsc{aic} values to judge which model is significantly better.
Therefore, there is a genuine need to develop a formal test that furnishes a critical value for nonnested model selection.

There exist many works on hypothesis testing for nonnested model selection. Cox (1961, 1962) pioneered a likelihood ratio (\textsc{lr}) test
for two separate families of hypotheses and heuristically argued its asymptotic normality, which was rigorously proven by White (1982a) under regularity conditions.
In a seminal article, Vuong (1989) used the Kullback-Leibler information criterion (\textsc{klic}) to measure the closeness of a model to the true data generating process (DGP). He introduced an \textsc{lr} test for competing models, which may be nested or non-nested and correctly specified or misspecified.
 However, it works only for parametric models
with known distributions.
Rivers and Voung (2002) extended this approach by replacing the likelihood with general lack of fit criteria, which allows for more estimation approaches,
but it still works only for parametric dynamic models.
Chen, Hong and Shum (2007) advanced a nonparametric \textsc{lr} test
for comparing a parametric likelihood model with a parametric moment condition model, based on the \textsc{klic} criterion, which can be regarded as extensions to the \textsc{lr} test of Vuong (1989).
McElroy (2016) proposed a Whittle \textsc{lr} test for nonnested model selection. This approach also employs the \textsc{klic} criterion and is designed for comparing two parametric
time series models with spectral densities.

However, the above \textsc{klic}-criterion based tests have different limiting distributions,
depending on whether the two models are overlapping or not,
and whether one of the model is correctly specified or not.
They require one to pretest which distribution to use before applying them.
As a consequence, they are basically two-step test procedures, which may induce the nonuniformity phenomenon of tests (Leeb and P\"otscher, 2005)
and result in size distortions (Shi, 2015; Schennach and Wilhelm, 2017).
In order to deal with this problem,
Shi (2015) proposed a one-step nondegenerated test for nonnested models, which is a modification of the Vuong test,
and Schennach and Wilhelm (2017) suggested a reweighted \textsc{lr} test for nonnested model selection. Both of the tests achieve uniform size control, but they are tailored for parametric models with densities.

Some authors introduced nonparametric extensions to the \textsc{lr} test.
Fan et al. (2001) proposed generalized likelihood ratio (\textsc{glr}) tests and showed that the Wilks type of results hold for a variety of useful models, including univariate non-parametric models, varying-coefficient models, and their extensions.
Fan and Jiang (2005) developed the \textsc{glr} test for  additive models based on the local polynomial fitting
 and the backfitting algorithm.
Fan et al. (2001), Fan and Huang (2005), and
 Fan and Jiang (2005, 2007) showed the generality of the Wilks phenomenon and enriched the applicability of the \textsc{glr} tests. However, the \textsc{glr} tests work only for nested models, require the working models contains the DGP,
and generally assume homogeneity of variance.
Moreover, the asymptotic distributionbs of \textsc{glr} tests explicitly depend on the bandwidth.
It remains unknown if the \textsc{glr} test can be modified for nonnested model selection.
Liao and Shi (2020) proposed a nondegenerate Vuong test
to comparison of nonnested nonparametric models, which employs
sieve approximations for M-estimation of the models, but the test requires correcting two bias terms and estimating the complicate variance, which explicitly depends on the tuning parameter in the sieve approximation.
In addition, it cannot deal with heteroscedastic errors, because their assumption 4.1(a) and the natrure of their M-estimate in eq.(3.3)  assume the error has a constant variance
 (see also Example~\ref{ex3}).

Last but not the least, it is worthy of mentioning that there are various metrics for model comparison within the Bayesian framework.
Two popular approaches among them are Bayes factors
(Lewis and Raftery, 1997)
and the deviance information criterion  (Spiegelhalter et al., 2002).
However, these methods are designed only for comparison of parametric models.

In this paper, we propose a general nonparametric test approach to model selection.
It is known that the prediction error criterion allows one to compare any two supervised statistical learning methods (parametric or nonparametric).
In practice, a statistical learning procedure with a smaller average (absolute or squared) prediction error ($APE$) is usually preferred. However, if the $APE$s are close among competing models,
one does not know if the $APE$s are significantly different.
Furthermore,
a model with a smaller $APE$ may be caused by
randomness of the sample, but not because of a better model.
These problems have hovered around statisticians over decades.
In an effort to solve them and to perform an accurate model selection,
we will resort to the idea of Owen (1988, 1989)
and propose some new \textsc{elr} tests
to compare prediction errors of competing models, based on the cross-validation method.
The proposed tests possess the following appealing characteristics:

{\spacingset{1.3}
\begin{itemize}
\item[(i)]
They are nonparametric tests without requiring a specific parametric structure or likelihood.
\item[(ii)]
The \textsc{elr} tests allow for heteroscedasticity of the errors,
and their asymptotic distributions and power do not depend on the smoothing parameters.
\item[(iii)] It allows one to fast implement the tests.
\item[(iv)] The tests have power to detect all $\sqrt{n}$ local alternatives.
\item[(v)]
 The idea is applicable to comparison between any two supervising statistical learning models, nested, non-nested, overlapping, correctly specified, or misspecified.
\end{itemize}
}

Because of the above features, the proposed \textsc{elr} test is robust against
heteroskedasticity, a striking contrast to the \textsc{glr} tests,
and it can be applied to post model inference (Tibshirani et al. 2016), for example, comparison between two post \textsc{lasso} models, nested or nonnested.
The \textsc{elr} test targets at comparing forecast equivalence of nonnested models, so can it be used to measure importance of explanatory variables for forecast in big data settings where mere significance tests do not make much sense (see Section~\ref{sec:bigdata}).

The empirical likelihood has been demonstrated as a powerful nonparametric tool for
interval estimates (Owen, 2001).
The  method has many advantages over the normal
approximation-based method and the bootstrap method
for constructing confidence intervals,
 such as the
transformation respecting, the range of parameter preserving,
the Bartlett correctable property, no requirement for estimating scale and skewness,
and no predetermined shape requirement
(Hall and La Scala, 1990).
There exists a vast literature devoted to the empirical likelihood for parametric models, but relatively less work for nonparametric models.
For interval estimation and hypothesis testing based on the empirical likelihood, they include but not limited to
Hall (1990),
Fan and Zhang (2004),
Chen and Keilegom (2009), etc.
However,
all these works formulate the \textsc{elr}
with some moment constraints from the estimation equations for the parameter of interest,
and no one acts for nonnested models.
In the construction of the proposed \textsc{elr} tests for nonnested models, we do not use
moment constraints from the estimation equations.
Existing techniques for the \textsc{elr} tests work only for correctly specified models and cannot be used to derive
the asymptotic distributions of the proposed \textsc{elr} tests.
These endow our work with challenges and
intelligence.
Since the \textsc{elr} tests employ the leave-one-out cross-validation (LOOCV) to calculate the prediction errors,
it is computationally expensive if one directly fits the models to the data with each
observation held out.
Due to the nature of global polynomial spline smoother used for competing models,
we are able to introduce a fast computation procedure
for implementation of the \textsc{elr} tests. This procedure requires us to fit the models to the data only once.
Furthermore, it is extended to test signficance of variables in additive models with massive or distributed data, and a distributed \textsc{elr} test is developed and posesses the same performance as the ideal \textsc{elr} test in the sense that there is no limited memory constraint and full data can be run on one machine.

This article is organized  as follows. In Section~\ref{secRPM} we describe the methodology.
The asymptotic distributions of our \textsc{elr} statistic are established whether or not the models are nested or misspecified,
from which a decision rule of model selection is proposed.
The fast implementation of the test is also considered.
In Section~\ref{sec:bigdata},  we develop the distributed \textsc{elr}
test for massive data.
In Section~\ref{sec:sim} we investigate finite sample
performance of \textsc{elr} tests via simulation, and in Section~\ref{sec:real} we provide an example of \textsc{elr} test on a real dataset.
Conditions and technical proofs are delegated to the Appendix.
\section{Methods}\label{secRPM}

Our main objective is to develop the \textsc{elr} theory for model selection.
To expose our idea, we consider model comparison between
the additive model and the varying coefficient model.
For other model comparison problems, our procedure can still applied but needs to be studied on a case-by-case basis.


\subsection{Model comparison based on prediction errors}\label{sec2}

Nonlinearity relationship exists widely in statistical theory and practice.
Suppose we have a random sample $\{y_{i},\bX_{i}, z_i\}_{i=1}^{n}$,
where
$\bX_i=(x_{i,1},\ldots,x_{i,p})^\top$,
 and we have found that there exists some in-sample significant evidence of ``nonlinearity" between $y_{i}$ and $\bX_{i}.$ We are
interested in further investigating whether the documented ``nonlinearity" is the true nonlinearity between $y_i$ and $\bX_i$,
or  is due to the functional coefficients in a linear regression model.
To deal with this problem,
we conduct model selection between the functional coefficient model
(Hastie and Tibshirani, 1993; Fan and Zhang, 1999; Cai, Fan and Yao, 2000)
\begin{equation}
y_i=\beta_0(z_i)+\sum_{j=1}^px_{i,j}\beta_j(z_i)+u_i, \label{3.1}
\end{equation}
and the nonparametric additive model (Hastie and Tibshirani, 1990)
\begin{equation}
y_i=\alpha+\sum_{j=1}^p m_j(x_{i,j})+v_i,\label{3.2}
\end{equation}
in the framework that both models may be wrongly specified,
where $z_i$ may be a component of $X_i$ or not,
and for identifiability it is assumed that
$E\{m_{j}(x_{i,j})\}=0$ .
Obviously, models (\ref{3.1}) and (\ref{3.2}) are nonnested in general,
but nested when $p=1$ and $z_i=x_{i,1}$.
They also overlap at the region where $y_i$ and $\bX_i$ are linearly related.

To get the prediction errors, we first need to estimate the unknown functions of
models (\ref{3.1}) and (\ref{3.2}).
Various estimation methods can be applied, such as
 the kernel smoother (Opsomer and Ruppert, 1997, 1998; Mammen and Park, 2006),
 the spline method (Stone, 1986; Zhou, Shen and Wolfe, 1998; Huang and Shen, 2004; Li and Liang, 2008),
and even the boosting learning algorithms (Freund and Schapire, 1997; Friedman, 2001).
In fact, one regards an estimation approach for a given supervising statistical model as a learning algorithm, and our \textsc{elr} test can compare any two
learning algorithms that provide predictions.
Then we need a good measure for assessing the performance of models (\ref{3.1}) and (\ref{3.2}).

A natural one is the prediction error from the widely used $K$-fold cross validation (CV) (Hastie and Tibshirani, 1990), even though other measures may be used.
This method randomly partitions the data into $K$ roughly equal-sized parts.
For the $k$th part, one uses the other $K-1$ parts of the data for training
and calculates the prediction error of each fitted model
when predicting the $k$th part of the data.
As in Hastie et al. (2009), we let $\theta:\{1,\ldots, n\} \to  \{1,\ldots, K\}$ be an indexing
function that indicates the partition to which observation $i$ is allocated by
the randomization, and let $\hat\alpha^{[-k]}$,
$\hat{\beta}_j^{[-k]}(\cdot)$
and
$\hat{m}_j^{[-k]}(\cdot)$
be fitted functions, computed with the $k$th part of the data removed.
In particular, when $K=n$, $\theta(i)=i$,
which corresponds to the leave-one-out CV (LOOCV).
Denote by
 $\hat{\varepsilon}_{1,i}= y_i-\hat{\beta}_0^{[-\theta(i)]}(z_i)-\sum_{j=1}^px_{i,j}\hat{\beta}_j^{[-\theta(i)]}(z_{i})$
and
$\hat{\varepsilon}_{2,i}=y_i-\hat{\alpha}^{[-\theta(i)]}-\sum_{j=1}^p \hat{m}_j^{[-\theta(i)]}(x_{i,j})$
the prediction errors for model (\ref{3.1}) and (\ref{3.2}), respectively.
Then the average (squared) prediction errors ($APE$) are
$$APE_1=n^{-1}\sum_{i=1}^n \hat{\varepsilon}_{1,i}^2\,\,\,
\mbox{\rm and}\,\,\,
APE_2=n^{-1}\sum_{i=1}^n \hat{\varepsilon}_{2,i}^2$$
for model (\ref{3.1}) and (\ref{3.2}), respectively.
Let $\hat{\xi}_{i}=\hat{\varepsilon}_{1,i}^2-\hat{\varepsilon}_{2,i}^2$. Then the difference
$$APE_1-APE_2=n^{-1}\sum_{i=1}^n\hat{\xi}_i$$
 is an appropriate estimate of
the difference of mean squared prediction errors
$$\mu_{\xi}=E(\hat{\xi}_i)=E(\hat{\varepsilon}_{1,i}^2)
-E(\hat{\varepsilon}_{2,i}^2)$$
between the two models,
and it
can be used to compare the performance of the two models
in terms of prediction.
When it is significantly different from zero, it signals that the two models are not competing. Otherwise, it is an indication of forecast equivalence.

\subsection{The ELR test}

Most existing model selection methods employ the likelihood or  information
criteria to measure the distance of a working model to the DGP.
However, for nonparametric models, the generalized likelihood ratio  method
works only for nested models and has some disadvantages,
and the information approach \textsc{klic} is not applicable,
as discussed in Section~\ref{sec1}.
We here introduce an \textsc{el} approach
to evaluate probability of forecast equivalence of two competing models.

As a nonparametric method, the \textsc{el} (Owen, 2001)
has become a standard approach to construct interval estimates.
 To use the \textsc{el}, one must specify estimating equations for the parameters of interest, but it is not necessary to estimate the variances of  the estimators of parameters. The latter property endows the \textsc{el} with
ability of handling heteroscedastic and asymmetric errors.
Note that
the nonparametric likelihood for forecast equivalence of the competing models is characterized by
$$\sup\{\prod_{i=1}^n p_i: p_i\ge 0, \sum_i p_i=1,\sum_i p_i\hat\xi_i=0\}.$$
Following the idea of (Owen, 1988, 1990; Qin and Lawless, 1994),
the above likelihood can be compared with nonparametric likelihood
 of a saturated model without any constraints,
in which all $p_i$ are equal to $1/n$.
Hence, we define  the logarithm of the \textsc{elr}
\begin{equation}\label{elik1}
R_{n,1}=-2\log\sup\bigl\{\prod_{i=1}^{n}(np_i):\, p\in \mathcal{G}\bigr\},
\end{equation}
where
$\mathcal{G}= \{p:\, p_i\geqslant 0, \sum_ip_i=1,
\sum_i p_i \hat\xi_i=0 \}.$
Note that
$\min\hat\xi_i\le \sum_i p_i \hat\xi_i\le \max \hat\xi_i.$
If $0\notin [\min\hat\xi_i, \max \hat\xi_i]$,
then $\mathcal{G}$ is empty and we set $R_{n,1}=+\infty.$

In the above construction, we do not set any
moment constraints from the estimation equations for both models (\ref{3.1}) and (\ref{3.2}). This is remarkably different from the classical  \textsc{elr} statistics where some moment constraints from the estimation equations are placed.
Using the Lagrange multiplier technique,
when $\min\hat\xi_i\le 0\le \max \hat\xi_i$,
we obtain that $p_i=n^{-1}\frac{1}{1+\lambda\hat\xi_i},$
 where $\lambda$ satisfies  that
\begin{equation}\label{eq3}
f(\lambda)\equiv \sum_{i=1}^{n}\hat\xi_{i}/(1+\lambda\hat\xi_i)=0.
\end{equation}
Let $\hat\lambda$ be the solution of equation (\ref{eq3}).
Then the logarithm of the \textsc{elr} becomes
\begin{equation}\label{eq4}
R_{n,1}=2\sum_{i=1}^{n}\log(1+\hat\lambda\hat\xi_{i}).
\end{equation}

If $\mu_{\xi}=0$, then the two models have equivalent performance in the sense that they have same prediction error on average,
that is, forecast equivalence (McElroy, 2016).
 The empirical likelihood ratio $R_{n,1}$ can be used to assess which model is better in terms of $APE$.
In fact, if the two models perform equivalently,
$R_{n,1}$ will be like a chi-squared random variable;
if one model is significantly better than the other,
then $R_{n,1}$ will go to infinity (see Theorem~\ref{th1} below).

Given the competing models (\ref{3.1}) and (\ref{3.2}),
one usually selects the model with a smaller $APE$,
but it is still unknown if the selected model is significantly better.
In other words, we need to develop a test to distinguish
if $APE_1-APE_2$ is significantly different from zero.
Therefore, we consider the following hypothesis testing problem:
\begin{equation}\label{hypo}
H_0^{(1)}: \mu_\xi= 0\,\
\mbox{\rm against}\,\
H_a^{(1)}: \mu_\xi\neq 0.
\end{equation}
The null $H_0$
means that models (\ref{3.1}) and (\ref{3.2}) perform equivalently according to prediction,
and the alternative represents one model is sufficiently better than the other.

\subsection{Asymptotic distributions and the decision rule}\label{sec2.3}


The $K$-fold CV is easy to implement, but for a given sample one has to select a value of $K$.
For ease of notations and for convenience of technical arguments, we only consider the  LOOCV. Even though our results hold for a general $K$-fold CV, but it requires $K$ to depend on sample size $n$ and  involves complicated specification of the rate of $K$ going to $\infty$ as $n\to\infty$, because theoretically the $K$-fold CV provides an asymptotic unbiased prediction only when $K\to\infty.$

For fitting models (\ref{3.1}) and (\ref{3.2}),
we need a smoothing method. Different smoothers can be employed,
and examples include the local linear smoother (Fan and Zhang, 1999; Fan and Jiang, 2005)
and the global polynomial spline smoothing (Stone,
1986; Li and Liang, 2008;  Jiang and Jiang, 2011), among other.
For illustration, we consider only the global polynomial spline smoothing,
which has stable performance and allows one to fast implement in the LOOCV (see the next section).


For model~\eqref{3.2}, we estimate $\alpha$ by $\bar{y}=n^{-1}\sum_{i=1}^ny_i$ and use B-spline basis functions to approximate each $m_{j}(\cdot)$.
Without loss of generality,
assume $\bX=(x_{1},\ldots,x_{p})^\top$ takes values in $\mathcal{W}=[0,1]^p$. For approximating function $m_{j}(\cdot)$, we need a knot sequence $\bar\phi_{j}=\{\phi_{j,k}\}_{k=0}^{q_j+1}$ such that
$0=\phi_{j,0}<\phi_{j,1}<\cdots<\phi_{j,q_j+1}=1$.
Denoted by
$\mathcal{S}(\ell_j,\bar\phi_j)$ the space of polynomial splines of order $\ell_j$ and knot sequence $\bar\phi_{j}$.
 Since $\mathcal{S}(\ell_j,\bar\phi_j)$ is a $\kappa_j$-dimensional linear space with $\kappa_j=q_j+\ell_j$,
for any $m_{j}\in \mathcal{S}(\ell_j,\bar\phi_j)$, there exists a local basis
$\{B_{j,k}(\cdot)\}_{k=1}^{\kappa_j}$
for $\mathcal{S}(\ell_j,\bar\phi_j)$, such that
$m_{j}(x_j)=\sum_{k=1}^{\kappa_j} b_{jk}B_{j,k}(x_j)$
 for $j=1,\ldots,p$
(Schumaker, 1981; Jiang and Jiang, 2011).
The local basis $\{B_{j,k}(\cdot)\}_{k=1}^{\kappa_j}$
depends on the knot sequence $\bar\phi_{j}$ and order $\ell_j$.
Let
$\bb_{j}=(b_{j1},\ldots,b_{j\kappa_j})^\top$,
$\bb=(\bb_1^\top,\ldots,\bb_p^\top)^\top$,
$\bPi_j(x_j)=(B_{j,1}(x_j),\ldots,B_{j,\kappa_j}(x_j))^\top,$
 and
 $\bPi(\bX)=(\bPi_1^\top(x_1),\ldots,\bPi_p^\top(x_p))^\top$.
For simplicity, denoted by $\mathrm{y}_i=y_i-\bar y$.
For any
$1\leq i\leq n$, we minimize the approximated sum of squared errors
\begin{equation}\label{bbs1}
\sum_{j=1(\neq i) }^n
\{\mathrm{y}_j-\bPi(\bX_j)^\top\bb\}^2
\end{equation}
over $\bb$,
which leads to the minimizer
 \begin{equation}\label{sufb}
 \hat\bb^{(-i)}=\bigl\{\bSigma_{n}^{(-i)}\bigr\}^{-1}
\bigl\{\frac{1}{n-1}\sum_{j=1(\neq i)}^{n}\bPi(\bX_j)\mathrm{y}_j\bigr\},
\end{equation}
where $\bSigma_{n}^{(-i)}=\frac{1}{n-1}\sum_{j=1(\neq i) }^n
  \bPi(\bX_j)\bPi(\bX_j)^\top.$
  Let
$
\hat{m}(\bX_{i})=\bPi(\bX_i)^\top\hat\bb^{(-i)}.
$
Then the prediction error  of model~\eqref{3.2} is given by
\begin{eqnarray}
  \hat\varepsilon_{2,i}=\mathrm{y}_i-\hat{m}(\bX_{i}).\label{APE2}
\end{eqnarray}

For functional coefficient model~(\ref{3.1}), we also assume $z_i$ takes values in $[0,1]$.
Similarly, there exits a local basis $\{B_{j,k}(\cdot)\}_{k=1}^{\widetilde\kappa_j}$ such that $\beta_{j}(z)=\sum_{k=1}^{\widetilde\kappa_j}B_{j,k}(z)c_{jk}$ for $j=0,1,\ldots,p$.
Let $\bc_{j}=(c_{j1},\ldots,c_{j\widetilde\kappa_j})^\top$, $\bc=(\bc_0^\top,\bc_1^\top,\ldots,\bc_p^\top)^\top$,
$\bGamma_j(z)=(B_{j,1}(z),\ldots,B_{j,\widetilde\kappa_j}(z))^\top,$
 and
 $\bGamma(\bX,z)=(\bGamma_0^\top(z),x_{1}\bGamma_1^\top(z),\ldots,x_p\bGamma_p^\top(z))^\top.$
For any
$1\leq i\leq n$, we minimize
\begin{eqnarray*}
\sum_{j=1(\neq i) }^n
\{y_j-\bGamma(\bX_j,z_j)^\top\bc\}^2
\end{eqnarray*}
over $\bc$ and get the minimizer
 $\hat\bc^{(-i)}=\bigl\{\bG_{n}^{(-i)}\bigr\}^{-1}
\bigl\{\frac{1}{n-1}\sum_{j=1(\neq i)}^{n-1}\bGamma(\bX_j,z_j)y_j\bigr\},$
where $\bG_{n}^{(-i)}=\frac{1}{n-1}\sum_{j=1(\neq i) }^n
  \bGamma(\bX_j,z_j)\bGamma(\bX_j,z_j)^\top$.
Then the prediction error of model~(\ref{3.1}) is
\begin{eqnarray}
  \hat\varepsilon_{1,i}=y_i-\bGamma(\bX_i,z_i)^\top\hat\bc^{(-i)}.
\label{APE1}
\end{eqnarray}


Given
 $\hat{\xi}_{i}=\hat{\varepsilon}_{1,i}^2-\hat{\varepsilon}_{2,i}^2$,
 we can calculate the \textsc{elr} statistic $R_{n,1}$ in \eqref{eq4}.
The following theorem describes its asymptotic null distribution.

\begin{theorem}\label{th1}
{\rm Assume that conditions A1 - A3 in Appendix A hold. Under $H_0^{(1)}$,
$ R_{n,1}\rightarrow \chi^2_1$ in distribution,
where
$\chi^2_1$ is the chi-squared distribution with one degree of freedom.
}\end{theorem}

%

Let $\chi_{1,1-\alpha}^2$ be the $(1-\alpha)$th percentile of $\chi^2_1$.
By  Theorem~\ref{th1}, at significance level $\alpha$, the rejection region of the \textsc{elr} test  is
$W=\{R_{n,1}> \chi_{1,1-\alpha}^2\}.$
To investigate the power of the proposed test, we consider the contiguous alternative of form:
\begin{equation}\label{hypoab}
H_{a,n}^{(1)}: \mu_{\xi}=a_n\sigma_{\xi} n^{-1/2},
\end{equation}
where $ \mu_{\xi}=E(\hat\xi_1)$,
$\sigma_{\xi}$ is the standard deviation of $\hat{\xi}_1$
and greater than zero,
and
$a_n$ is a sequence of real numbers such that $\lim_{n\to\infty} a_n=a$.
Then the power of the \textsc{elr} test can be approximated using the following theorem.

\begin{theorem}\label{th1a}
{\rm Suppose conditions A1 - A3 hold. Under $H_{a,n}^{(1)}$,
$R_{n,1}\rightarrow \chi^2_1(a^2)$ if $|a|<+\infty$,
and
$P( R_{n,1}\to +\infty)\to 1$ if $|a|=+\infty$,
where
$\chi^2_1(a^2)$ is the noncentral chi-squared distribution with one degree of freedom
and noncentral parameter $a^2$.}
\end{theorem}
%

Theorems~\ref{th1}-\ref{th1a} have an interesting implication. We can approximate the power of the test by
$$P_{H_{a,n}^{(1)}}(W)\approx  P(\chi^2_1(a^2)> \chi_{1,1-\alpha}^2)
=1-\{\Phi(|a|+\sqrt{\chi_{1,1-\alpha}^2})-\Phi(|a|-\sqrt{\chi_{1,1-\alpha}^2})\},$$
where $\Phi(\cdot)$ is the distribution function of ${\mathcal N}(0,1).$
The power function is increasing in $|a|$
and shares the same formula as that of the likelihood ratio test
for testing $H_0:\, \mu=0$ against $H_{1n}:\,\mu=a_n\sigma n^{-1/2}$,
based on an iid sample of size $n$ from
the normal population ${\mathcal N}(\mu,\sigma^2).$
This suggests that the proposed test is powerful.

From Theorems \ref{th1}-\ref{th1a}, given a significant level $\alpha$, we conduct a model selection procedure based on the following decision rule:
{\spacingset{1.3}
\begin{itemize}
\item[(i)] If $R_{n,1}<\chi_{1,1-\alpha}^2$, then we cannot reject $H_0^{(1)}: \mu_{\xi} = 0$, and we say the two models are asymptotically equivalent.
\item[(ii)]
If $R_{n,1}>\chi_{1,1-\alpha}^2$, one model is sufficiently better than the other.
Furthermore,
\begin{itemize}
\item[(a)]
 if  $APE_1<APE_2$, model (\ref{3.1}) is better than model (\ref{3.2});
\item[(b)]
if  $APE_1>APE_2$, model (\ref{3.2}) is better than model (\ref{3.1}).
\end{itemize}
\end{itemize}
}

Our \textsc{elr} test is asymptotically chi-squared under $H_0$ that the two models are forecast equivalent, no matter if the models are nested, nonnested, overlapping, correctly specified, or misspecified.
As shown in Theorem~\ref{th1a}, it has nontrivial power against
all local alternatives $H_{a,n}^{(1)}$ which converge to the null at rate
$\sqrt{n}$ or faster ($|a|\le +\infty$).
Unlike Vuong's type of tests,
we do not pretest which distribution to use for calculating the critical value, and thus
the \textsc{elr} test can uniformly control the size of test as in Shi (2015) and
Schennach and Wilhelm (2017).
We test the forecast equivalence against non-equivalence
of the two models.
If the null is rejected, we retain the model with smaller $APE$;
otherwise, we believe both models provide equal forecast performance.
In any cases, we make this kind of conclusions, no matter if the models are nested or not and misspecified or not,
which is consistent with the framework of likelihood inference under model misspecification
in White (1982b).

\subsection{Fast implementation}\label{sec2.4}

The \textsc{elr} test involves the LOOCV for calculating
the prediction errors, which requires fitting the models
to each subsample with one observation held out.
In the following we introduce a fast algorithm for computing the prediction errors.

Define the projection matrices
$\bP_{A}=\bD(\bD^\top\bD)^{-1}\bD^\top$
and
$\bP_{V}= \bE(\bE^\top\bE)^{-1}\bE^\top,$
where $\bD=(\bPi(\bX_1), \ldots,\bPi(\bX_n))^\top$ and $\bE=(\bGamma(\bX_1,z_1),\ldots,\bGamma(\bX_n,z_n))^\top$.
Let the residual vectors be
 $\be_{1}=(e_{1,1},\ldots,e_{1,n})^\top=\bY-\bP_{V}\bY$ and
$\be_{2}=(e_{2,1},\ldots,e_{2,n})^\top=\mathrm{Y}-\bP_{A}\mathrm{Y},$
where $\bY=(y_1,\ldots,y_n)$ and $\mathrm{Y}=(\mathrm{y}_{1},\ldots,\mathrm{y}_{n})^\top$.
Then, using the classical technique in linear models, we obtain that
$$
\hat{\bb}^{(-i)}=\hat{\bb}
-(1-p_{A,i})^{-1} (\bD^\top\bD)^{-1}\bPi(\bX_i)e_{2,i},
$$
$$\hat{\bc}^{(-i)}=\hat{\bc}
-(1-p_{V,i})^{-1} (\bE^\top\bE)^{-1}\bGamma(\bX_i,z_i)e_{1,i},$$
where
 $\hat\bb=(\bD^\top\bD)^{-1}\bD^\top\mathrm{Y},$
$\hat\bc=(\bE^\top\bE)^{-1}\bE^\top\bY,$
 and
 $p_{A,i}$ and $p_{V,i}$ are the $i$th diagonal entries of the matrices $\bP_{A}$ and $\bP_{V}$, respectively.
Furthermore,
\begin{equation}\label{eqs1}
\hat\varepsilon_{1,i}=(1-p_{V,i})^{-1}e_{1,i}
\,\,\mbox{\rm and}\,\,
\hat\varepsilon_{2,i}=(1-p_{A,i})^{-1}e_{2,i}.
\end{equation}
Hence,   $\hat{\xi}_{i}=\hat{\varepsilon}_{1,i}^2-\hat{\varepsilon}_{2,i}^2$ can be calculated by fitting the models to full data only.
Since $f'(\lambda)=\sum_{i=1}^n \hat{\xi}_i^2/(1+\lambda\hat{\xi}_i)^2<0,$ $f(\lambda)$ is strictly decreasing.
 Then
 evaluation of the \textsc{elr} statistic $R_{n,1}$ is straightforward
while solving equation \eqref{eq3} to obtain $\hat\lambda$ by the Newton-Raphson iterations with initial value $\lambda=0$.

\section{Application to big data}\label{sec:bigdata}

In most cases the sample size of big data is huge,
and existing statistical inference methods face up to challenges.
Consider fitting a linear model with big data, for example,
the p value of a t-statistic for an individual coefficient
is possibly less than $5\%$. Since no model is right, the p value goes to zero as sample
size $n$ goes to $\infty$, no matter how small the coefficient is.
Even if the model is correct, the p value may be very small for a nonzero coefficient as sample size $n$ gets large enough.

However, for a very small coefficient the corresponding
covariate may not be of practical importance at all. That is, statistical
significance may not imply practical importance in a big data setting. Naturally, one may ask how to measure practical importance of a covariate  if there is a huge sample. In other words,
we need some
measures to calibrate the importance of explanatory variables (or their functional forms), rather than merely assessing their statistical significance.
This is expected to be a challenge in statistical analysis for massive data where the memory of one machine cannot fit all the data.

\subsection{Distributed ELR test}

As we discussed before,
no model is right,
and any statistical model can be misspecified in practice.
Therefore, it will make much more sense to make comparison between
misspecified models than concentrating on statistical significance of covariates in big data settings. This is particularly relevant to economic modeling, because it is possible that more than one economic
models, some of which can be even conflicting to each other, coexist in
explaining the same economic phenomenon, and the existing econometric tools cannot distinguish them from each other for various reasons.
Obviously, our \textsc{elr} test can be used for this task, and in particular it can be used to compare the two models with or without  an explanatory variable.

 Consider modeling a massive dataset, for example, using the additive model (\ref{3.2}).
To evaluate importance of the $\ell$th variable for forecast, we compare model (\ref{3.2}) with
\begin{equation}
y_i=\alpha+\sum_{j=1(\neq \ell)}^p m_j(x_{i,j})+v_i, \,\,\, i=1,\ldots,n.\label{3.2a1}
\end{equation}
Let $\bPi_{-\ell}(\bX)=(\bPi_1^\top(x_1),\ldots,\bPi_{\ell-1}^\top(x_{\ell-1}),\bPi_{\ell+1}^\top(x_{\ell+1}),\ldots,\bPi_p^\top(x_p))^\top$
and
$$\bSigma^{(-i)}_{n,-\ell}=\frac{1}{n-1}\sum_{j=1(\neq i) }^n
  \bPi_{-\ell}(\bX_j)\bPi_{-\ell}(\bX_j)^\top.$$
Then, similar to \eqref{sufb}, the spline coefficient $\bb$ is estimated by
 \begin{equation}\label{eqj115}
 \hat\bb^{(-i)}_{-\ell}=\bigl\{\bSigma_{n,-\ell}^{(-i)}\bigr\}^{-1}
\bigl\{\frac{1}{n-1}\sum_{j=1(\neq i) }^n\bPi_{-\ell}(\bX_j)\mathrm{y}_j\bigr\}.
\end{equation}
Similar to \eqref{APE2}, we obtain the prediction errors from model \eqref{3.2a1}:
\begin{eqnarray}\label{APE3}
 \hat\varepsilon_{3,i}
 =\mathrm{y}_i-\bPi_{-\ell}(\bX_i)^\top\hat\bb^{(-i)}_{-\ell}.
\end{eqnarray}
Then the difference of squared prediction errors between model \eqref{3.2a1} and \eqref{3.2}
are
 given by
 $\hat{\eta}_i=\hat\varepsilon^{2}_{3,i}-\hat\varepsilon^{2}_{2,i},$
 which can be calculated quickly
 using the same technique as in \eqref{eqs1}.
 Similar to \eqref{hypo}, comparing model  (\ref{3.2}) to model~\eqref{3.2a1} reduces to testing
 \begin{equation}\label{hypo2}
H_0^{(2)}: \mu_\eta= 0\,\
\mbox{\rm against}\,\
H_a^{(2)}: \mu_\eta\neq 0,
\end{equation}
 where
 $\mu_\eta=E(\hat{\eta_1}).$
 Using the same argument as for $R_{n,1}$, we obtain the \textsc{elr} statistic
\begin{equation}\label{eq4a1}
R_{n,2}=2\sum_{i=1}^{n}\log(1+\hat\nu\hat\eta_i),
\end{equation}
where $\hat\nu$ satisfies that
$\sum_{i=1}^{n}\hat\eta_{i}/(1+\hat\nu\hat\eta_i)=0.$
As argued for $R_{n,1}$, if $0\notin [\min\hat\eta_i, \max \hat\eta_i]$,
we set $R_{n,2}=+\infty.$

Large values suggest rejection of $H_0^{(2)}.$
If $H_0^{(2)}$ is rejected, then it suggests that the $\ell$th covariate is practically important for forecasting the response.
This is a variable selection problem in which
both models are nested but may be misspecified.
Existing approaches deal with it by assuming the larger model is correctly specified. However, our \textsc{elr} test does not require this condition.

Some challenges arise when we use $R_{n,2}$ for massive or distributed data. Practically, we need to solve the computation problem
since the classical computation methods for estimating $m_j$'s
 and
for empirical likelihood ratio (Hall and La Scala, 1990) are computationally infeasible for massive data.
We need to develop some distributed computing methods to solve this problem.
The existing divide-and-conquer method (Zhang et al., 2013; Chen and Xie, 2014; Chen et al., 2019; Battey et al., 2018) can be employed in general, but the resulting limiting distribution of the test statistic should be consistent with that of the original test with full data, or some other inference methods such as the bootstrap procedure adaptive massive data (Chen and Peng, 2018) are to be advanced.
In the following we will work on these problems
and provide a distributed \textsc{elr} test for nonnested model selection with massive data.
Remarkably, our distributed test will perform the same as the original test.

Suppose we have a massive sample of size $n=Nm$. Then we randomly split the entire dataset $\{\mathrm{y}_i,\bX_i, 1\leq i\leq n\}$ into $N$ subsamples $\mD_1,\ldots,\mD_N$, each of which has the same size $m=n/N$.
For distributed data, the full sample consists of these subsamples
installed on $N$ machines at different sites.
If different machines have different subsample sizes,
our procedure can be straightforwardly extended.
Typically, using the divide-and-conquer algorithm one fits the models with each subsample on each machine
and gets the prediction error for each subsample point on each machine, and integrates them to form the \textsc{elr} test. Since each prediction error uses information from only a subsample, the resulting \textsc{elr} test will not be as powerful as the original \textsc{elr} test with full data.
Even if one calculates the prediction error with full sample information,
the resulting \textsc{elr} test will not have the same finite sample performance as the original \textsc{elr} test.
 Instead of fitting the models to the subsample on every machine, we calculate
only some sufficient statistics from each subsample
and use them to estimate the spline coefficients.
Then the estimated coefficients are feedback to each slaver
so that the LOOCV error can be calculated.

Specifically, let $\mD_k=\{\mathrm{y}_j^{(k)},\bX^{(k)}_j,j=1,\ldots,m\}$
be the subsample distributed on the $k$th machine  for $1\leq k\leq N$.
Then there exists a one to one mapping $\nu:\, \{1,\ldots,m\}\otimes \{1,\ldots,N\}\to \{1,\ldots,n\}$ such that
\begin{equation}\label{18a}
i=\nu(j,k)\,\,\
\mbox{\rm and}\,\,\
 (\mathrm{y}^{(k)}_{j},\bX_{j}^{(k)})=(\mathrm{y}_i,\bX_i)
 \,\,\ \mbox{\rm for }\,\,\ i=1,\ldots,n.
 \end{equation}
For example, $\nu(j,k)=j+(k-1)m$ is such a mapping.

Note that
the B-spline basis vector $\bPi(\bX_{i})$ depends on the knot sequences $\{\bar\phi_{j}\}_{j=1}^p$ with $\bar\phi_{j}=\{\phi_{j,k}\}_{k=0}^{q_j+1}$.
For each subset $\mD_k$, we can compute $\{\bPi(\bX^{(k)}_{j}), j=1,\ldots,m\}$ and $\{\bPi_{-\ell}(\bX^{(k)}_{j}), j=1,\ldots,m\}$ with some given knot sequences $\{\bar\phi_{j}\}_{j=1}^p$ independent of $k$.
Choice of such knot sequences for massive or distributed data will be discussed in Section~\ref{3.2a}.
It follows from \eqref{18a} that, for $\alpha=0,\ell$,
\begin{eqnarray*}\label{Dbspline}
\bPi_{-\alpha}(\bX^{(k)}_{j})=\bPi_{-\alpha}(\bX_{i}),
\end{eqnarray*}
where, with a little abuse of notations, we denote $\bPi(\bX_{i})$ by
$\bPi_{-0}(\bX^{(k)}_{j})$ for convenience.
Note that the LOOCV estimators for models \eqref{3.2} and \eqref{3.2a1} involve only statistics:
$$\mathcal{X}_{\alpha}^{(-i)}\equiv\sum_{s=1(\neq i)}^{n}\bPi_{-\alpha}(\bX_{s})\bPi_{-\alpha}(\bX_{s})^\top
\,\,\
\mbox{\rm and}\,\,\
\mathfrak{F}_{\alpha}^{(-i)}\equiv\sum_{s=1(\neq i)}^{n}\bPi_{-\alpha}(\bX_{s})\mathrm{y}_{s}$$
for $i=1,\ldots,n$.
Let
$$\mathcal{A}_{\alpha}=\sum_{k=1}^{N}A_{-\alpha}^{(k)}
\,\,\ \mbox{\rm and}\,\,\
 \mathcal{B}_{\alpha}=\sum_{k=1}^{N}B^{(k)}_{-\alpha}, $$
where
$A_{-\alpha}^{(k)}=\sum_{s=1}^{m}\bPi_{-\alpha}(\bX^{(k)}_{s})\bPi_{-\alpha}(\bX^{(k)}_{s})^\top$
and
$B^{(k)}_{-\alpha}=\sum_{s=1}^{m}\bPi_{-\alpha}(\bX^{(k)}_{s})\mathrm{y}_{s}^{(k)}$
are sufficient statistics for the subsample on the $k$th machine.
These sufficient statistics can be calculated on individual machines.
Then
\begin{eqnarray*}
\mathcal{X}_{\alpha}^{(-i)}
= \mathcal{A}_{\alpha}-\bPi_{-\alpha}(\bX^{(k)}_{j})\bPi_{-\alpha}(\bX^{(k)}_{j})^\top\,\,\
\mbox{\rm and}\,\,\
\mathfrak{F}_{\alpha}^{(-i)}
= \mathcal{B}_{\alpha}-\bPi_{-\alpha}(\bX^{(k)}_{j})\mathrm{y}^{(k)}_{j}.
\end{eqnarray*}
Hence,
 the LOOCV estimators in \eqref{sufb} and \eqref{eqj115} are  rewritten as
 $$\hat\bb^{(k)}_{\alpha,j}=\bigl\{ \mathcal{A}_{\alpha}-\bPi_{-\alpha}(\bX^{(k)}_{j})\bPi_{-\alpha}(\bX^{(k)}_{j})^\top\bigr\}^{-1}
\bigl\{\mathcal{B}_{\alpha}-\bPi_{-\alpha}(\bX^{(k)}_{j})\mathrm{y}^{(k)}_{j}\bigr\},$$
respectively for $\alpha=0,\ell$.
Then distributed prediction errors from
 model~\eqref{3.2} and \eqref{3.2a1} are given by
\begin{eqnarray*}\label{APE2a}
\hat\varepsilon^{(k)}_{2,j}=\mathrm{y}^{(k)}_j-\bPi_{-0}(\bX^{(k)}_j)^\top \hat\bb_{0,j}^{(k)}\,\,\
\mbox{\rm and}\,\,\
 \hat\varepsilon^{(k)}_{3,j}
 = \mathrm{y}_j^{(k)}-\bPi_{-\ell}(\bX^{(k)}_j)^\top\hat\bb_{\ell,j}^{(k)} ,
\end{eqnarray*}
 respectively.
Therefore, for $\mathrm{y}_j^{(k)}$ the difference of squared prediction errors from models \eqref{3.2a1} and \eqref{3.2}
 is given by
 $\hat{\eta}_j^{(k)}=|\hat\varepsilon^{(k)}_{3,j}|^2-|\hat\varepsilon^{(k)}_{2,j}|^2$.
Then, similar to \eqref{eq4a1}, we obtain the distributed \textsc{elr} statistic
\begin{equation}\label{eq4a2}
R_{n,3}=2\sum_{k=1}^{N}\sum_{j=1}^{m}\log(1+\hat\tau\hat\eta^{(k)}_j),
\end{equation}
where $\hat\tau$ satisfies that
$\sum_{k=1}^{N}\sum_{j=1}^{m}\hat\eta^{(k)}_{j}/(1+\hat\tau\hat\eta^{(k)}_j)=0.$
Again,
if $0\notin [\min\hat\eta^{(k)}_j, \max \hat\eta^{(k)}_j]$,
we set $R_{n,3}=+\infty.$
The root $\hat{\tau}$ of the above equation can be found via the distributed Newton-Raphson iterations,
that is, be implemented on each machine.

Since $\hat\eta_{j}^{(k)}=\hat\eta_{i}$, we have
$R_{n,2}=R_{n,3}$.
That is, they have the same finite-sample and asymptotic performance,
and distributed test $R_{n,3}$ has the same power as  ideal test $R_{n,2}$ with no memory constraint.
The following theorem depicts the asymptotic null distributions of
the \textsc{elr} tests.

\begin{theorem}\label{th2}
{\rm Suppose conditions A2 and A4 hold. Then, under $H_0^{(2)}$,
 $R_{n,3}\rightarrow \chi^2_1$ in distribution.
}\end{theorem}

To study the power of test,
we consider testing $H_{0}^{(2)}$ against a sequence of contiguous alternatives:
\begin{equation}\label{hypo1}
H^{(2)}_{a,n}: \mu_{\eta}=a_n\sigma_{\eta}n^{-1/2},
\end{equation}
where $\mu_{\eta}=E(\hat\eta_{1})$, $\sigma_{\eta}$ is the standard deviation of $\hat\eta_1$ and greater than zero,
and
$a_n$ is the same as in \eqref{hypoab}.
In the following we present the alternative distributions of the \textsc{elr} tests.

\begin{theorem}\label{th2a}
{\rm Assume that conditions A2 and A4 hold. Then,
under $H^{(2)}_{a,n}$,
$R_{n,3}\rightarrow \chi^2_1(a^2)$
if $|a|<+\infty$,
and
$P(R_{n,3}\to +\infty)\to 1$ if $|a|=+\infty$.
}\end{theorem}

From Theorems \ref{th2}-\ref{th2a}, given a significant level $\alpha$, we conduct a variable selection procedure based on the following decision rule:
{\spacingset{1.3}
\begin{itemize}
\item[(i)] If $R_{n,3}<\chi_{1,1-\alpha}^2$, then we cannot reject $H_0^{(2)}$. According to Occam's razor, we choose model~\eqref{3.2a1} without the $\ell$th covariate.
\item[(ii)]
If $R_{n,3}>\chi_{1,1-\alpha}^2$, it suggests the $\ell$th covariate is practically important for forecasting the response.
We choose model~\eqref{3.2} as the working model.
\end{itemize}
}

In the above decision rule,
we choose model~\eqref{3.2} when the null is rejected.  This agrees with choosing the working model close to the true.
Since model~\eqref{3.2} is larger,
it is closer to the true than model~\eqref{3.2a1}.

\subsection{Knot selection with massive data }\label{3.2a}
There are two popular ways of deciding the knots. One is to place equally spaced knot sequence, and the other is to use the
 quantile knot sequence
from the empirical distribution of the underlying variable.
The 1st knot choice can be easily computed since it is independent of the data. For the 2nd knot choice,
it seems that the sample quantiles of the x-variables for massive or distributed data are not easy to get,
but the median-searching algorithm in Harris (2012) can be adapted to the current situation. Specifically,
we consider how to get the $q$th quantile of  the sample $\{x_{i,1}, i=1,\ldots, n\}$
 for any $q\in (0,1)$.
Let $x_{(1),1}\leq x_{(2),1}\leq \cdots \leq x_{(n),1}$ be the order statistics. Then, the $q$th sample quantile is defined by
\begin{equation}\label{quantile}
x_{(\lfloor h\rfloor),1}+(h-\lfloor h\rfloor)\{x_{(\lceil h\rceil),1}-x_{(\lfloor h\rfloor),1}\},
\end{equation}
 where $\lfloor h\rfloor$(or $\lceil h\rceil$) denotes the nearest integer to $h\equiv (n-1)q+1$, which is less (or larger) than $h$. This is the default way of defining sample quantile in software R, and is equivalent to the Excel and Python optional ``inclusive'' methods.

 According to \eqref{quantile},
 it suffices to find
  $x_{(l),1}$ for any $1\leq l\leq n$.
 Let us split
 sample $\{x_{i,1}\}_{i=1}^n$
 into $N$ sets $\mathcal{F}_{k}=\{x_{j,1}^{(k)},\,j=1,\ldots,m\}$ for $1\leq k\leq N$.
Then our distributed algorithm proceeds as follows:
{\spacingset{1.3}
\begin{itemize}
  \item [(i)] Randomly select a set $\mathcal{F}_{k}$ and an element $a\in\mathcal{F}_{k}$;
  \item [(ii)] Compute subset $\mathcal{C}_{k'}=\{x\in \mathcal{F}_{k'}:\, x\leq a\}$ for $k'=1,\ldots,N$
  and
  $\mathcal{N}=\sum_{k'=1}^{N}|\mathcal{C}_{k'}|$ with $|\mathcal{C}_{k'}|$ being the number of elements of $\mathcal{C}_{k'}$;
  \item [(iii)] If $\mathcal{N}=l$, the algorithm stops and returns  $a$
  as  $x_{(l),1}$;
  \item [(iv)] If $\mathcal{N}>l$, renew $\mathcal{F}_{k'}=\mathcal{C}_{k'}$ for $k'=1,\ldots,N$ and go to step (i);
  \item [(v)] If $\mathcal{N}<l$, renew $\mathcal{F}_{k'}=\mathcal{F}_{k'}/\mathcal{C}_{k'}$ for $k'=1,\ldots,N$  and $l=l-\mathcal{N}$,
   and go to step (i).
\end{itemize}
}
As mentioned in Harris (2012), the computational complexity of the above algorithm is $O\{n/N + N\log(n/N)\}$.
When $N\ll \sqrt{n}$, the computational complexity is simply $O(n/N)$,
which decreases as $N$ increases.
 For the split-and-conquer method, it usually assumes the technical condition $N\ll \sqrt{n}$, but our method works for any $1\le N\le n$.

{\spacingset{1.5}
\begin{algorithm}[htbp]\caption{ Distributed \textsc{elr} algorithm }\label{A1}
\footnotesize
\KwIn{$\{y_i,\bX_i, 1\leq i\leq n\}$, $n$, $m$, $N$, $\tau_0=0$, $\{\bar\phi_{j}\}_{j=1}^p$,  $\omega$ and $\varphi$;}
\KwOut{$R_{n,3}$; }

{\bf Initialization:} Randomly partition $\{\mathrm{y}_i,\bX_i, 1\leq i\leq n\}$ into $N$ subsets $\{\mathrm{y}^{(k)}_j,\bX^{(k)}_j,\,j=1,\ldots,m\}$ for $1\leq k\leq N$ and distribute them on $N$ machines\;
{\bf Circulation:}
      \For{$k=1:N$}
        { { With the knot sequences $\{\bar\phi_{j}\}_{j=1}^p$,
         compute B-spline basis
         $\{\bPi(\bX^{(k)}_{s}),\, s=1,\ldots,m\}$
         and
         $\{\bPi_{-\ell}(\bX^{(k)}_{s}), \,s=1,\ldots,m\}$}\;
          {$A_{-0}^{(k)}=\sum_{s=1}^{m}\bPi(\bX^{(k)}_{s})\bPi(\bX^{(k)}_{s})^\top$
        and $A^{(k)}_{-\ell}=\sum_{s=1}^{m}\bPi_{-\ell}(\bX^{(k)}_{s})\bPi_{-\ell}(\bX^{(k)}_{s})^\top$}\;
          {$B^{(k)}_{-0}=\sum_{s=1}^{m}\bPi(\bX^{(k)}_{s})\mathrm{y}_{s}^{(k)}$ and
            $B^{(k)}_{-\ell}=\sum_{s=1}^{m}\bPi_{-\ell}(\bX^{(k)}_{s})\mathrm{y}_{s}^{(k)}$ }\;
        }
 For $\alpha=0,\ell$,
 compute $\mathcal{A}_{\alpha}=\sum_{k=1}^{N}A_{-\alpha}^{(k)}$, $\mathcal{B}_{\alpha}=\sum_{k=1}^{N}B^{(k)}_{-\alpha}$,
         $\hat\bb_{-\alpha}=\mathcal{A}_{\alpha}^{-1}\mathcal{B}_{\alpha}$;

     {\bf Circulation:} \For{$k=1:N, j=1:m$}
            {  
             {$e_{2,j}^{(k)}=\mathrm{y}^{(k)}_j-\bPi(\bX^{(k)}_j)^\top\hat\bb_{-0}$,
             $e_{3,j}^{(k)}=\mathrm{y}^{(k)}_j-\bPi_{-\ell}(\bX^{(k)}_j)^\top\hat\bb_{-\ell}$
             }\;
             {$p_{0,j}^{(k)}=\bPi(\bX^{(k)}_j)^\top\mathcal{A}_{0}^{-1}\bPi(\bX^{(k)}_j)$,
             $p_{\ell,j}^{(k)}=\bPi_{-\ell}(\bX^{(k)}_j)^\top\mathcal{A}_{\ell}^{-1}\bPi_{-\ell}(\bX^{(k)}_j)$
             }\;
                 {$\hat\varepsilon^{(k)}_{2,j}=(1-p_{0,j}^{(k)})^{-1}e_{2,i}^{(k)}$,
                 $\hat\varepsilon^{(k)}_{3,j}=(1-p_{\ell,j}^{(k)})^{-1}e_{3,j}^{(k)}$
                 }\;

                 $\hat\eta_{j}^{(k)}=|\hat\varepsilon^{(k)}_{3,j}|^2-|\hat\varepsilon^{(k)}_{2,j}|^{2}$;

            }
\uIf{$\min_{j,k}\hat\eta_{j}^{(k)}\leq 0\leq \max_{j,k}\hat\eta_{j}^{(k)}$}{

{\bf Iteration:} \For{$t=1:\omega$}{
 {\bf Circulation:} \For{$k=1:N$}
{$D_{1k}=\sum_{j=1}^{m}\frac{\hat\eta^{(k)}_{j}}{1+\tau_{t-1}\hat\eta_j^{(k)}},$
$D_{2k}= \sum_{j=1}^{m}\frac{|\hat\eta^{(k)}_{j}|^{2}}{|1+\tau_{t-1}\hat\eta_j^{(k)}|^2}$;
}
  {$D_1=\sum_{k=1}^{N}D_{1k}$,
   $D_2=-\sum_{k=1}^{N}D_{2k}$},
 $\tau_{t}=\tau_{t-1}-D_1/D_2$\;
 \If{$|\tau_{t}-\tau_{t-1}|<\varphi$}{
 $\hat\tau=\tau_{t}$\;
  {\bf Circulation:} \For{$k=1:N$}
  {$R_{n,3}^{(k)}=2\sum_{i=1}^{m}\log(1+\hat\tau\hat\eta^{(k)}_{i})$}
 $R_{n,3}=\sum_{k=1}^{N}R_{n,3}^{(k)}$;
}
}

}
\Else{
$R_{n,3}=10^{16};$
}

{\bf Return} $R_{n,3}$.
\end{algorithm}
}

\subsection{A distributed ELR algorithm}

Computational details of the distributed \textsc{elr} algorithm is listed in Algorithm~\ref{A1}.
Note that
 the full sample estimators of coefficient vector $\bb$
 for models  \eqref{3.2} and \eqref{3.2a1} are
 $\hat\bb_{-\alpha}=\mathcal{A}_{\alpha}^{-1}\mathcal{B}_{\alpha}$,
 respectively for $\alpha=0,\ell$. Let $e_{2,j}^{(k)}=\mathrm{y}^{(k)}_j-\bPi(\bX^{(k)}_j)^\top\hat\bb_{-0}$ and
 $e_{3,j}^{(k)}=\mathrm{y}^{(k)}_j-\bPi_{-\ell}(\bX^{(k)}_j)^\top\hat\bb_{-\ell}$.
 Then, similar to \eqref{eqs1},
 we have the following fast calculation formulas for the LOOCV prediction errors:
  $$\hat\varepsilon^{(k)}_{2,j}=(1-p_{0,j}^{(k)})^{-1}e_{2,j}^{(k)}\ \ \,
  \mbox{\rm and}\,\,\
       \hat\varepsilon^{(k)}_{3,j}=(1-p_{\ell,j}^{(k)})^{-1}e_{3,j}^{(k)},$$
  where
  $p_{\alpha,j}^{(k)}=\bPi_{-\alpha}(\bX^{(k)}_j)^\top\mathcal{A}_{\alpha}^{-1}\bPi_{-\alpha}(\bX^{(k)}_j)$
  for $\alpha=0,\ell$.
The above formulas have been incorporated into Algorithm~\ref{A1}.
In this algorithm, we calculate the sufficient statistics $A_{-\alpha}^{(k)}$
and $B_{-\alpha}^{(k)}$ on individual machines,
with which we obtain
 $\mathcal{A}_{\alpha}$ and  $\mathcal{B}_{\alpha}$. Then
 the full sample coefficient estimator $\hat\bb_{-\alpha}$
and the LOOCV prediction errors are calculated.
At last, we evaluate test statistic $R_{n,3}$.

\section{Simulations}\label{sec:sim}

To investigate the size and power of our \textsc{elr} test for nonnested models,
 we conduct simulations for model selection  
 in different situations.
For each of the following examples, we run 600 simulations,
 and for each simulation we generated an iid sample
 from the DGP.  For each simulation,
the cubic B-splines were used to estimate the unknown functions in the working models,
and the leave-one-out CV method was employed to calculate the $APE$.
The number of knots is chosen by an adjusted $APE$ criterion.
Specifically,
for model~\eqref{3.2}
 $\{\kappa_{j}, j=1,\ldots,p\}$
 were chosen by minimizing the adjusted $APE$
$$APE_{adj}= \frac{1}{n-\kappa}\sum_{i=1}^n|\hat\varepsilon_{2,i}|^{2},$$
where $\kappa=\sum_{j=1}^p\kappa_{j}$.
For models~\eqref{3.1} and~\eqref{3.2a1}, the $APE_{adj}$ is defined similarly but with
$(\hat\varepsilon_{2,i},\kappa)$ replaced by $(\hat\varepsilon_{1,i},\tilde\kappa)$ and $(\hat\varepsilon_{3,i},\kappa-\kappa_{\ell})$, respectively.

In Example \ref{ex2}, we investigate if the proposed \textsc{elr} test works for mis-specified nonnested models with heteroscedasticity.
In Example \ref{ex3}, we compare our test with the \textsc{glr} test in Fan and Jiang (2005, JASA) and  the uniform Vuong (\textsc{unv}) test in Liao and Shi (2020), we also study robustness of these tests.
 In Example~\ref{ex4}, we consider our distributed \textsc{elr} test for massive data.


\begin{example}\label{ex2}{\rm
Consider model selection between
varying-coefficient model
$y_{i}=\beta_0(z_{i})+\beta_1(z_{i})x_{i,1}+\beta_2(z_{i})x_{i,2}+u_{i}$
and
additive model
$y_{i}=\alpha+m_{1}(x_{i,1})+m_{2}(x_{i,2})+v_{i}$,
with iid samples generated from
\begin{eqnarray}\label{sim2}
y_i&=&0.5(x_{i,1}+x_{i,2})+\theta\{x_{i,1}\exp(1+z_{i})+x_{i,2}1(z_i>0.5)+1.5\cos(\pi z_{i})\}\nonumber\\
&&+\tau\{\exp(x_{i,1})\cos( x_{i,1})+0.5\sin(x_{i,2})\}+\sin(\pi x_{i,1})\varepsilon_i,
\end{eqnarray}
where
$(x_{i,1}, x_{i,2})$ are bivariate normally distributed with standard normal marginals and correlation coefficient $0.5$,
 $z_i\sim U(0,1)$,
 and
$\varepsilon_i\sim N(0,1).$

The varying coefficient models and the additive model are nonnested.
When $\theta=\tau=0$, both models are correctly specified;
when $\theta=0$ and $\tau\neq 0$,
the additive model is correctly specified;
when $\theta\neq 0$ and $\tau=0$, the varying coefficient model is correctly specified;
when $\theta\neq 0$ and $\tau\neq 0$, both models are misspecified.
We set different values of $\theta$ and $\tau$
to evaluate the size and power of our test.
Since the GDP has changing variance,
it allows us to evaluate the performance of our \textsc{elr} test
when the error is heteroscedastic.
{\spacingset{1.3}
\begin{table}[htbp]
\centering\small
\caption{Null rejection rates (\%) of \textsc{elr} tests
at significance level 5\%(left cell) and 10\%(right cell)
for Example~\ref{ex2}}\label{tab2}
\begin{tabular}{|c| c|  c| c | c|  c | c|} \hline
n & \multicolumn{6}{c}{$(\theta,\tau)$}\vline\\
\cline{2-7}
 & \multicolumn{1}{c}{(0, 0)}  & \multicolumn{1}{c}{(0,0.07)} &\multicolumn{1}{c}{ (0, 0.09)} & \multicolumn{1}{c}{(0, 0.12)} & \multicolumn{1}{c}{(0, 0.15)} & \multicolumn{1}{c}{(0, 0.18)} \vline\\\hline
1000 & (4.00,9.00)  & (12.3,16.5)  &(27.0,34.8)  & (55.7,65.8) & (83.8,86.8)& (94.0,95.8)\\
1500 &(5.33,9.33)  & (38.3,44.7)   &(63.5,70.8)   &(87.0,91.2) & (96.5,97.8) &(99.2,100) \\
\hline
Model
selection &\multicolumn{1}{c}{Both}\vline &\multicolumn{1}{c}{Additive}\vline &\multicolumn{1}{c}{Additive}\vline &\multicolumn{1}{c}{Additive} \vline&\multicolumn{1}{c}{Additive} \vline&\multicolumn{1}{c}{Additive}  \vline  \\ \hline
Sign of
DAPE & &\multicolumn{1}{c}{+}\vline &\multicolumn{1}{c}{+}\vline &\multicolumn{1}{c}{+} \vline&\multicolumn{1}{c}{+} \vline&\multicolumn{1}{c}{+}  \vline  \\
\hline
n & \multicolumn{6}{c}{$(\theta,\tau)$}\vline\\
\cline{2-7}
 & \multicolumn{1}{c}{(0, 0)}  & \multicolumn{1}{c}{(0.05, 0)} &\multicolumn{1}{c}{ (0.075, 0)} & \multicolumn{1}{c}{(0.1, 0)} & \multicolumn{1}{c}{(0.125, 0)} & \multicolumn{1}{c}{(0.15, 0)} \vline\\\hline
1000  & (4.00,9.00)   & (45.8,58.3) & (68.5,80.7) & (88.7,94.5)&  (96.7,98.3)&  (99.0,99.3)\\
1500 &(5.33,9.33) & (57.8,71.0)  &(80.2,87.7)  &(96.0,97.8)  & (99.5,99.7) &(100,100) \\
\hline
Model
selection &\multicolumn{1}{c}{Both}\vline &\multicolumn{1}{c}{Varying}\vline &\multicolumn{1}{c}{Varying}\vline &\multicolumn{1}{c}{Varying} \vline&\multicolumn{1}{c}{Varying} \vline&\multicolumn{1}{c}{Varying}  \vline  \\ \hline
Sign of
DAPE & &\multicolumn{1}{c}{$-$}\vline &\multicolumn{1}{c}{$-$}\vline &\multicolumn{1}{c}{$-$} \vline&\multicolumn{1}{c}{$-$} \vline&\multicolumn{1}{c}{$-$}  \vline  \\
\hline
n & \multicolumn{6}{c}{$(\theta,\tau)$}\vline\\
\cline{2-7}
 & \multicolumn{1}{c}{(0.05, 0.05)}  & \multicolumn{1}{c}{(0.18, 0.1)} &\multicolumn{1}{c}{ (0.18,0.05)} & \multicolumn{1}{c}{(0.05, 0.18)} & \multicolumn{1}{c}{(0.1, 0.18)} & \multicolumn{1}{c}{(0.18, 0.18)} \vline\\\hline
1000 & (13.3,22.3) &(96.5,98.5)  & (99.7,100)  & (64.5,74.3)& (16.3,24.0)   &(29.8,39.5)\\
1500 & (8.33,14,5) & (99.7,99.8)   & (100,100)  & (92.5,95.3)&  (36.3,46.2)&(26.5,34.8)\\
\hline
Model
selection &\multicolumn{1}{c}{Varying}\vline &\multicolumn{1}{c}{Varying}\vline &\multicolumn{1}{c}{Varying}\vline &\multicolumn{1}{c}{Additive} \vline&\multicolumn{1}{c}{Additive} \vline&\multicolumn{1}{c}{Varying}  \vline  \\ \hline
Sign of
DAPE & \multicolumn{1}{c}{$-$}\vline&\multicolumn{1}{c}{$-$}\vline &\multicolumn{1}{c}{$-$}\vline &\multicolumn{1}{c}{$+$} \vline&\multicolumn{1}{c}{$+$} \vline&\multicolumn{1}{c}{$-$}  \vline  \\
\hline
\end{tabular}
{DAPE -  average of the differences of APEs between model~\eqref{3.1} and model \eqref{3.2} in 600 simulations }
\end{table}
}

For each paired values of $(\theta,\tau)$,
we calculated the null rejection rates of our \textsc{elr} tests for testing problem (\ref{hypo})
at $5\%$ and $10\%$  significance levels.
The simulation results are summarized in Table~\ref{tab2}.
It is seen that
our \textsc{elr} test uniformly controls size over
different significance levels,
since the reject rates are all close to the nominal size at $(\theta,\tau)=(0,0)$.
When one of $\theta$ and $\tau$ goes far away from $0$ and the other is fixed at $0$, the alternative runs further away from the null, and
the rejection rate becomes higher and higher,
which reveals that our test gets more and more powerful.
When both $\theta$ and $\tau$ are nonzero,
the two models are nonnested and misspecified, and
the power gets higher as the distance between $\theta$ and $\tau$ increases.
This demonstrates that our \textsc{elr} test  works great here.
$\hfill\diamond$
}\end{example}

{\spacingset{1.3}
\begin{table}[htbp]
\centering\small
\caption{Null rejection rates (\%) of the \textsc{elr}, \textsc{unv},
 and \textsc{glr} tests
 for Example~\ref{ex3}}\label{tab3}
\begin{tabular}{|c|c|c|c |  c|  c | c|  c | c|} \hline
DGP& Test  &n &\multicolumn{6}{c}{$\tau$}\vline\\
\cline{4-9}
& &  &\multicolumn{1}{c}{0} \vline   &\multicolumn{1}{c}{0.06}\vline  & \multicolumn{1}{c}{0.08}\vline & \multicolumn{1}{c}{0.1} \vline & \multicolumn{1}{c}{0.12}\vline  & \multicolumn{1}{c}{0.16} \vline\\ \hline
\multirow{6}{*}{normal} &\multirow{2}{*}{ELR} &1000 & 5.17  & 16.5  &39.0   & 78.3 & 93.3& 97.3\\
 & &1500 &4.50 & 26.0   &64.0   &94.7  & 98.3 &99.2 \\ \cline{2-9}
 &\multirow{2}{*}{UNV} &1000 & 4.17  & 23.0  &49.3   & 83.8& 93.8& 98.0\\
 & &1500 &3.83 & 34.0   &71.8   &96.2  & 99.3 &99.8 \\\cline{2-9}
  &\multirow{2}{*}{GLR} &1000 & 5.83  & 20.8  &45.5   & 79.5 & 90.7& 96.7\\
 & &1500 &5.00 & 34.7   &73.7   &93.2  & 99.0 &99.7 \\
\hline
\multirow{6}{*}{conditionally normal} &\multirow{2}{*}{ELR} &1000 & 5.17  & 45.2   & 79.0  & 96.8& 98.3& 99.2\\
 & &1500 &6.50  & 71.2   &95.5  &98.5 & 99.0&99.7 \\ \cline{2-9}
&\multirow{2}{*}{UNV} &1000 & 2.80  & 61.7  &88.5 & 98.2  & 99.3 & 99.5\\
 & &1500 &1.33 & 84.5   &99.0   &99.8  & 100 &100 \\\cline{2-9}
  &\multirow{2}{*}{GLR} &1000 & 14.8  & 62.8  &88.7   & 97.7 & 99.5& 100\\
 & &1500  & 13.3  & 79.5  &98.3   & 100 & 100& 100\\
\hline
\multirow{6}{*}{conditional t(6)} &\multirow{2}{*}{ELR}  &1000  & 5.17  & 23.5 &65.2 & 90.3  & 96.7&98.3 \\
 & &1500 &5.67  &46.8   &89.0 &98.2 &99.3  &100\\\cline{2-9}
&\multirow{2}{*}{UNV} &1000 & 2.00  & 32.3  &75.3   & 94.2 & 97.8& 98.5\\
 & &1500 &0.83 & 55.0   &93.0   &99.2  & 100 &100 \\\cline{2-9}
  &\multirow{2}{*}{GLR} &1000 & 13.8  & 40.0  &73.5   & 95.0 & 98.2& 99.8\\
 & &1500 &12.3 & 54.0   &91.2   &99.3  & 99.5 &100 \\
\hline
\multirow{6}{*}{mixed normal} &\multirow{2}{*}{ELR}&1000 & 5.83 & 12.5   & 27.0  & 62.5& 83.2& 93.2\\
 & &1500  &5.33  &17.2   &51.8 &88.8 &95.7  &98.3\\\cline{2-9}
&\multirow{2}{*}{UNV} &1000 & 2.00  & 12.5  &34.3   & 69.7 & 85.1& 94.5\\
 & &1500 &2.00 & 16.7   &53.0   &88.3  & 95.6 &98.8 \\\cline{2-9}
\  &\multirow{2}{*}{GLR} &1000 & 5.67  & 15.7  &32.0   & 60.5 & 81.0& 94.0\\
 & &1500 &5.17 & 20.2   &48.8   &82.7  & 94.2 &98.2 \\
\hline
\end{tabular}
\end{table}
}

\begin{example}\label{ex3}{\rm
 Let us consider
model comparison between
varying-coefficient model
$y_{i}=\beta_0(z_{i})+\beta_1(z_{i})x_{i,1}+u_{i}$
and additive model
$y_{i}=\alpha+m_{1}(x_{i,1})+m_{2}(x_{i,2})+v_{i},$
when the true DGP is
$$y_i=0.5x_{i,1}+0.25x_{i,1}\cos(x_{i,1})+\tau\exp(x_{i,2})\cos( x_{i,2})+\varepsilon_{i},
$$
where
$(x_{i,1}, x_{i,2})$ are the same as in Example~\ref{ex2},
and
 $\varepsilon_{i}$ is
$N(0,1)$ (normal),
$\sin(x_{i,2})N(0,1)$ (conditionally normal),
$\sin(x_{i,2})t(6)$ (conditional t(6)),
and
$0.95N(0,1)+0.05 N(0,3^2)$ (mixed normal), respectively.
This allows us to assess robustness of our \textsc{elr} test for model comparison under a variety of noises.
We set $z_i=x_{i,1}$ to make the additive model contains the varying coefficient model, so that the \textsc{glr} test can be applied.
These tests are also compared to the \textsc{unv} test.

Table~\ref{tab3} reports the powers of the three tests.
It is seen that the \textsc{elr} test not only keeps the size but also is nearly most powerful and  robust against the error distributions.
For the normal and mixed normal errors,
our \textsc{elr} test has nearly the same power as the \textsc{glr} test.
As expected, the \textsc{glr} and \textsc{unv} tests cannot keep the size when the errors are heteroscedastic.
$\hfill\diamond$

%
%
%
%
%
}\end{example}

\begin{example}\label{ex4}{\rm
Consider comparing
models
$y_{i}=\alpha+m_{1}(x_{i,1})+m_{2}(x_{i,2})+v_{i}$
and
$y_{i}=\alpha+m_{2}(x_{i,2})+v_{i}$ with massive data,
when the true DGP is
\begin{eqnarray}\label{sim4}
y_i=\tau\exp(x_{i,1})\cos(x_{i,1})+0.1x_{i,2}(1+x_{i,2})+\varepsilon_i,
\end{eqnarray}
where
$\varepsilon_i$ is $\sin(\pi x_{i,2}){\mathcal N}(0, 1)$
and
$(x_{i,1},x_{i,2})$ are the same as in Example~\ref{ex2}.
Obviously, this is a model with heteroscedascity.

{\spacingset{1.3}
\begin{table}[!htbp]
\centering
\centering\small
\caption{Null rejection rates (\%) of \textsc{elr} tests
for Example~\ref{ex4}}\label{tab4}
\begin{tabular}{|c|c|c | c|  c | c|  c | c|} \hline
N&m & \multicolumn{6}{c}{$\tau$}\vline\\
\cline{3-8}
 & & \multicolumn{1}{c}{0} \vline  & \multicolumn{1}{c}{0.01} \vline &\multicolumn{1}{c}{0.015}\vline  & \multicolumn{1}{c}{0.02}\vline  & \multicolumn{1}{c}{0.025}\vline  & \multicolumn{1}{c}{0.03} \vline\\\hline
\multirow{2}{*}{1}&21000 & 5.50 & 15.0 & 48.0 &90.2  & 98.3 &99.5 \\
&42000 &5.67 & 33.3 & 84.5 &99.0  &100  &100 \\
\hline
\multirow{2}{*}{50}&420 & 5.50 & 15.0 & 48.0 &90.2  & 98.3 &99.5 \\
&840 &5.67 & 33.3 & 84.5 &99.0  &100  &100 \\
\hline
\multirow{2}{*}{100}&210& 5.50 & 15.0 & 48.0 &90.2  & 98.3 &99.5 \\
&420 &5.67 & 33.3 & 84.5 &99.0  &100  &100 \\
\hline
\multirow{2}{*}{150}&140 & 5.50 & 15.0 & 48.0 &90.2  & 98.3 &99.5 \\
&280 &5.67 & 33.3 & 84.5 &99.0  &100  &100 \\
\hline
\end{tabular}
\end{table}
}

We set sample size $n=21,000$ and $42,000$
for comparing full sample test $R_{n,2}$ with distributed test $R_{n,3}$.
Table~\ref{tab4} reports the rejection rate of $H_0^{(2)}$ at 5\% significance level. When $\tau=0$,
 the null and the alternative coincide,
 and the power is the type I error probability.
 As shown in the table, all powers are close to the nominal level 5\% at $\tau=0$,
 indicating our tests keep the size.  As $\tau$ increases, the alternative moves further away from the null,
 and the rejection rate of the null gets higher and higher. Furthermore,
 for  different numbers of machines $N=1, 50,100,150$,
$R_{n,2}$ and $R_{n,3}$ have the same performance. This implies that our distributed  \textsc{elr} tests can exactly recovery the results of the original  \textsc{elr} test with the whole data running on one machine.

}\end{example}

%

\section{A real example}\label{sec:real}
We illustrate our method by analyzing the Boston housing dataset.
This dataset contains information collected by the U.S Census Service concerning housing in the area of Boston, MA. It is available at the StatLib archive ({{\url{http://lib.stat.cmu.edu/datasets
/boston}}}). The dataset consists of median values of owner-occupied homes in 506 homes and several variables that might explain the variation of housing value (Harrison and Rubinfeld, 1978; Fan and Huang, 2005). Fan and Huang (2005) considered the following seven variables: CRIM (per capita crime rate by town), RM (average number of rooms per dwelling), TAX (full-value property-tax rate per $\$$10,000), NOX (nitric oxides concentration parts per 10 million), PTRATIO (pupil-teacher ratio by town),
AGE (proportion of owner-occupied units built prior to 1940),
 and LSTAT (lower status of the population). For simplicity, the variables CRIM, RM, $\log(\text{TAX})$, NOX, PTRATIO and AGE are denoted by $x_{1}, x_{2},\ldots, x_{6}$, respectively.
 Let $y$ be the response (median value of owner-occupied homes)
 and
  $z=\log(\text{LSTAT})$.

  The object is to study the association between $y$ and $\bX=(x_{1}, x_{2},\ldots, x_{6})$, given a sample
  $\{y_i,\bX_i, z_i, i=1,\ldots,n\}$ with size $n=506$.
 Many authors analyzed the dataset using different models.
 Examples include the additive models in  Opsomer and Ruppert (1998) and Fan and Jiang (2005), and the varying coefficient model in Fan and Huang (2005), among others.
 However, there is no formal model comparison among them. In the following we use our \textsc{elr} test to do this work. In all cases, the significance level is taken as $5\%$.

\begin{itemize}
\item[(i)] (Varying coefficient model vs additive model)
 Fan and Huang (2005) considered the varying coefficent model:
\begin{equation}\label{eq3.1a1a}
E(y_i|z_i,\bX_i)=\beta_0(z_i)+\sum_{j=1}^6x_{i,j}\beta_j(z_i).
\end{equation}
 We are
interested in further investigating whether the documented ``nonlinearity" is the true nonlinearity between $y_i$ and $\bX_i$,
or  is due to the functional coefficients in a linear regression model.
Thus, we consider the following nonparametric additive model for comparison:
\begin{equation}\label{eq3.2a1a}
E(y_i-\bar y|z_i,\bX_i)=m_{0}(z_i)+\sum_{j=1}^6 m_j(x_{i,j}),
\end{equation}
with $\bar y=n^{-1}\sum_{i=1}^n y_i$,
which contains the models studied in  Opsomer and Ruppert (1998)
and Fan and Jiang (2005).
This reduces to model selection between models~\eqref{eq3.1a1a} and~\eqref{eq3.2a1a}. Based on the sample, the value of ELR statistic is $19.33$, greater than the critical vale $\chi_{1,0.95}^2=3.84$,
 and the average difference of squared prediction errors between  models \eqref{eq3.1a1a} and~\eqref{eq3.2a1a}
is given by $n^{-1}\sum_{i=1}^n\hat\xi_i=19.36$.
Hence, according to the decision rule below Theorem~\ref{th1a}, we choose model~\eqref{eq3.2a1a}.

\item[(ii)] (Comparison between additive models)
  Opsomer and Ruppert (1998) analyzed the dataset via a four dimensional additive model:
\begin{eqnarray}\label{3.2aa}
E(y_i-\bar y|z_i,\bX_i)=m_{0}(z_i)+m_2(x_{i,2})+m_{3}(x_{i,3})+m_{5}(x_{i,5}).
\end{eqnarray}
 Based on the \textsc{glr} test for the above model,
 Fan and Jiang (2005) confirmed to fit the dataset with the following semiparametric model:
   \begin{eqnarray}\label{3.3aa}
E(y_i-\bar y|z_i,\bX_i)=a_0z_i+m_2(x_{i,2})+a_3x_{i,3}+a_5x_{i,5}.
\end{eqnarray}

First, we consider model selection between models~\eqref{3.3aa} and~\eqref{3.2aa} using our \textsc{elr} test.
The realized value  of ELR statistic is $14.57$, which is greater than
the critical value $\chi_{1,0.95}^2$,
and
the average difference of squared prediction errors
between models~\eqref{3.3aa} and~\eqref{3.2aa}
is given by $n^{-1}\sum_{i=1}^n\hat\xi_i=2.67$. This suggests us to choose~\eqref{3.2aa}, which agrees with the \textsc{glr} test.

Next, we compare model~\eqref{3.2aa} with  model~\eqref{eq3.2a1a}. The \textsc{elr} statistic is $30.53>\chi_{1,0.95}^2$.
The average difference of squared prediction errors
is given by $n^{-1}\sum_{i=1}^n\hat\xi_i=4.39$. This leads to selection of model~\eqref{eq3.2a1a}.
That is, at least one of $m_{1}(\cdot)$, $m_{4}(\cdot)$ and
$m_{6}(\cdot)$ are not zero.

 Then, we test $H_{0\ell}:\, m_{\ell}(\cdot)=0$ against
 $H_{1\ell}:\, m_{\ell}(\cdot)\neq 0$
  for each $\ell=1,4,6$
   in model~\eqref{eq3.2a1a}, using the \textsc{elr} test.
  The results are reported in Table~\ref{R3}.
{\spacingset{1.3}
\begin{table}[htbp]
\centering\centering\small
\caption{ELR testing  whether a nonparamatric function is zero}
\label{R3}
\begin{tabular}{cccc} \hline
&$m_{1}(\cdot)$&$m_{4}(\cdot)$&$m_{6}(\cdot)$\\
\hline
ELR& 8.20 & 10.92& 0.82 \\
\hline
Equivalent(=)&$\neq$ &$\neq$ &$=$\\
\hline
\end{tabular}
\end{table}
}
Obviously, $m_1$ and $m_{4}$ are statistically significant, but $m_6(\cdot)$ not at $5\%$ significance level, based on individual \textsc{elr} tests or the multiple \textsc{elr} test with the Bonferroni correction.
This leads to the model
 \begin{eqnarray}\label{3.2aaa}
E(y_i-\bar y|z_i,\bX_i)=m_{0}(z_i)+m_{1}(x_{i,1})+m_2(x_{i,2})+m_{3}(x_{i,3})+m_{4}(x_{i,4})+m_{5}(x_{i,5}).
\end{eqnarray}

Last, we compare model~\eqref{3.2aaa} with model~\eqref{eq3.2a1a}. The ELR statistic is $0.82<\chi_{1,0.95}^2$.
Thus, models~\eqref{eq3.2a1a} and~\eqref{3.2aaa} are equivalent.
Since model \eqref{3.2aaa} is simpler, it is preferable according to the Occam's razor. This selection agrees to Table~\ref{R3}.
\end{itemize}





\begin{center}
{\bf\large Appendix}:  Notations and Conditions
\end{center}

 \newtheorem{Lemma}{Lemma}
\renewcommand{\theequation}{A.\arabic{equation}}
\setcounter{equation}{0}

For ease of exposure, we introduce some notations which will be used throughout the remainder of the paper.
 Let $\bSigma_{A}=E\{\bPi(\bX)\bPi(\bX)^\top\}$, $\bSigma_{A,-l}=E\{\bPi_{-\ell}(\bX)\bPi_{-\ell}(\bX)^\top\}$,
  and
$\bSigma_{C}=E\{\bGamma(\bX,z)\bGamma(\bX,z)^\top\}$.
Put
  $\widetilde\bPi(\bX_i)=\bSigma_{A}^{-1/2}\bPi(\bX_i)$,
  $\widetilde\bPi_{-\ell}(\bX_i)=\bSigma_{A,-l}^{-1/2}
\bPi_{-\ell}(\bX_i)$,
 and $\widetilde\bGamma(\bX_i,z_i)=\bSigma_{C}^{-1/2}\bGamma(\bX_i,z_i)$.
For $\alpha=0,-\ell$, let
$$\tilde\bb_{-\alpha}=\arg\min_{\bb}E\{\mathrm{y}-\bPi_{-\alpha}(\bX)^\top\bb\}^2\,\, \
\mbox{\rm and}\,\,\
\tilde\bc=\arg\min_{\bc}E\{y-\bGamma(\bX,z)^\top\bc\}^2.$$
Then, by the first order condition, the above minimizers admit closed formulas:
$$\tilde\bb_{-\alpha}=\{E(\bPi_{-\alpha}(\bX)\bPi_{-\alpha}(\bX)^\top)\}^{-1}E\{\bPi_{-\alpha}(\bX)\mathrm{y}\};\ \
\tilde\bc=\{E(\bGamma(\bX,z)\bGamma(\bX,z)^\top)\}^{-1}E\{\bGamma(\bX,z)y\}.
$$
The population versions of prediction errors for models \eqref{3.1}, \eqref{3.2} and \eqref{3.2a1}
are
$\tilde\varepsilon_{1,i}=y_i-\bGamma(\bX_i,z_i)^\top\tilde\bc$,
$\tilde\varepsilon_{2,i}=\mathrm{y}_i-\bPi(\bX_i)^\top\tilde\bb_{-0}$,
 and
$\tilde\varepsilon_{3,i}=\mathrm{y}_i-\bPi(\bX_i)_{-\ell}^\top\tilde\bb_{-\ell}$,
respectively.
It is straightforward to verify that
 \begin{equation}\label{8a}
\tilde\varepsilon_{1,i}=y_i-\widetilde\bGamma(\bX_i,z_i)^\top E\{\widetilde\bGamma(\bX_1,z_1)y_1\},
\end{equation}
\begin{equation}\label{9a}
\tilde\varepsilon_{2,i}=\mathrm{y}_i-\widetilde\bPi(\bX_i)^\top E\{\widetilde\bPi(\bX_1)\mathrm{y}_1\},
\end{equation}
\begin{equation}\label{10a}
\tilde\varepsilon_{3,i}=\mathrm{y}_i-\widetilde\bPi_{-\ell}(\bX_i)^\top E\{\widetilde\bPi_{-\ell}(\bX_1)\mathrm{y}_1\}.
\end{equation}
Then we define
$\tilde\xi_i=\tilde\varepsilon_{1,i}^2-\tilde\varepsilon_{2,i}^2$
 and $\tilde\eta_i=\tilde\varepsilon_{3,i}^2-\tilde\varepsilon_{2,i}^2$,
 which are population versions of $\hat{\xi}_i$ and $\hat{\eta}_i$,
 respectively.

To establish our theoretical results, we need some technical conditions.
Let
$\mH_r$ be a space of functions whose $d$th order derivative is
H$\ddot{o}$lder continuous of order $v$.
That is,
$\mH_r=\{h(\cdot):\, |h^{(d)}(a')-h^{(d)}(a)|\leq C|a'-a|^v,a,a'\in[0,1]\}$, where $h^{(d)}(\cdot)$ is $d$th derivative
and $r= d+v$.
If $v=1$, then $h^{(d)}(\cdot)$ is Lipschitz continuous.
Assume the following conditions hold:

{\spacingset{1.3}
\begin{itemize}\label{C1}
\item[A1]
(Varying coefficient model)
(i)  The eigenvalues of matrix $E\{\bGamma(\bX,z)\bGamma(\bX,z)^\top\}$ are bounded away from $0$ and $\infty$;
(ii) Assume that $\beta_{j}\in \mH_r$,
 and $\widetilde\kappa_j=O( n^{1/(2r+1)})$ for some $r> 1.5$ and $0\leq j\leq p$;
(iii) Assume there exists some $\gamma>2$ such that
$E\|\bGamma(\bX,z)\|_2^{2\gamma}=O(\widetilde\kappa^\gamma)$, 
and
$E|y|^{2\gamma}<+\infty$,
where $\widetilde\kappa=\sum_{j=0}^p\widetilde\kappa_{j}$.
\item[A2]
(Additive model)
(i)  The eigenvalues of matrix $E\{\bPi(\bX)\bPi(\bX)^\top\}$ are bounded away from $0$ and $\infty$;
(ii) Assume that $m_{j}(\cdot)\in \mH_r$
 and $\kappa_j=O( n^{1/(2r+1)})$ for $r>1.5$
and $1\leq j\leq p$;
(iii) Assume there exists some $\gamma>2$ such that
$E\|\bPi(\bX)\|_2^{2\gamma}=O(\kappa^\gamma),$
 and
 $E|y|^{2\gamma}<+\infty$,
where $\kappa=\sum_{j=1}^p\kappa_{j}$.

\item[A3]
(Varying coefficient and additive models) Assume that
$E|\tilde\varepsilon_{1,1}|^{2\gamma}=O(1), E|\tilde\varepsilon_{2,1}|^{2\gamma}=O(1)$,
 and $\text{Var}(\tilde\xi_1)> c_1$ for some constants $c_1>0$
and $\gamma>2$.

\item[A4]
(Additive model)
Assume that
$E|\tilde\varepsilon_{2,1}|^{2\gamma}=O(1),$
 $E|\tilde\varepsilon_{3,1}|^{2\gamma}=O(1)$, and
$\text{Var}(\tilde\eta_1)>c_1$ for some constants $c_1>0$
and $\gamma>2$.

\end{itemize}
}
The above conditions are wild.
 By Lemma 7 of Tang et al. (2013),
 condition A1(i) holds.
 Condition A2(i) is the same as condition A.2 of Belloni et al. (2015).
 Conditions A1(ii) and  A2(ii) were assumed in Theorem~1 of
Tang et al. (2013).
For B-spline series, Newey (1997) assumed $\sup_{x_{j}}\|\bPi_{j}(x_{j})\|_{2}=O(\sqrt{\kappa_{j}})$, which implies
our condition
$E\|\bPi(\bX)\|_{2}^{2\gamma}=O(\kappa^{\gamma})$
in A2(iii).
Notice that
$\bGamma_j(z)=(B_{j,1}(z),\ldots,B_{j,\widetilde\kappa_j}(z))^\top$
and
$\bGamma(\bX,z)=(\bGamma_0^\top(z),x_{1}\bGamma_1^\top(z),\ldots,x_p\bGamma_p^\top(z))^\top$.
If $E(|x_j|^{2\gamma}\mid z_j=z)$ is a bounded function of $z$
and
$E\|\bGamma_j(z)\|_{2}^{2\gamma}=O(\widetilde\kappa^{\gamma}),$
then the condition
$E\|\bGamma(\bX,z)\|_{2}^{2\gamma}=O(\widetilde\kappa^{\gamma})$
in A1(iii) holds.

Since the squared prediction errors in the \textsc{elr} test are compared,
conditions A3 and A4 assume that the $\gamma$th moments of their population versions
 must be bounded away from $+\infty$, i.e. $E(|\tilde\varepsilon_{j,1}|^{2\gamma})=O(1)$ for $j=1,2,3$.
This can be relaxed if one compares the median of prediction errors, but it will complicate the technical proofs of theorems.
Furthermore, it is assumed in condition A3 that $\text{Var}(\tilde\xi_1)> c_1$.
This condition, combined with Lemma~\ref{le4}(i), ensures that $\sigma_\xi>0$. Otherwise,
there is no need to develop a test for comparson of the two competing models.
Similarly, in condition A4 it is reasonable to assume $\text{Var}(\tilde\eta_1)>c_1$.
\noindent{\bf Supplementary Material}

To save space, all technical proofs of theorems are included in the online supplementary material.

{\spacingset{1.2}
\bibliographystyle{Chicago}
\bibliography{Bibliography-MM-MC}
}

\newpage
\setcounter{page}{1}

\bigskip
\begin{center}
{\Large\bf Supplementary material for ``Nonnested model selection
based on empirical likelihood''}
\end{center}



Now we give technical proofs of our theorems.
To streamline our arguments, we first introduce some technical lemmas
whose proofs are reported after the proofs of theorems.
%
%
%
%
%

\begin{Lemma}\label{le1}
\rm{
Assume conditions A1 - A3 hold. Then, for $j=1,2$,
\begin{itemize}
\item[(i)]
$\max_{1\leq i\leq n}|\tilde\varepsilon_{j,i}-\hat\varepsilon_{j,i}|=O_{P}(n^{\frac{1}{2\gamma}+\frac{3}{4r+2}-\frac{1}{2}})$;
(ii)
$n^{-1}\sum_{i=1}^{n}(\hat\varepsilon_{j,i}-\tilde\varepsilon_{j,i})^2=O_{P}(n^{\frac{2}{2r+1}-1})$.
\end{itemize}
}
\end{Lemma}


\begin{Lemma}\label{le2}
\rm{
Assume conditions A1 - A3 hold.  Then
\begin{itemize}
\item[(i)] $n^{-1}\sum_{i=1}^n(\hat\xi_i-\tilde\xi_i)=o_{P}(n^{-1/2})$
;
(ii)
$n^{-1}\sum_{i=1}^n (\hat\xi_i^2-\tilde\xi_i^2)=o_{P}(1)$.
\end{itemize}
}
\end{Lemma}


\begin{Lemma}\label{le4}
\rm{
Assume conditions A1 - A3 hold.  Then 
\begin{itemize}
\item[(i)]
$\text{\Var}(\tilde\xi_1)=\sigma_{\xi}^2+o(1)$;
\item[(ii)]
 $\max_{1\leq i\leq n}|\tilde\xi_i|=O_{P}(n^{1/\gamma})$, $\max_{1\leq i\leq n}|\hat\xi_i|=O_{P}(n^{1/\gamma})$,
$E|\tilde\xi_1|^{\gamma}=O(1)$,
 and
$n^{-1}\sum_{i=1}^n\tilde\xi_i^2=E\tilde\xi_1^2+o_{P}(1)$;
\item[(iii)]under $H^{(1)}_{a,n}$, $a_n^{-1}\sqrt{n}E\tilde\xi_1/\sigma_{\xi}\to 1$ when $|a|=+\infty$,
 and $\sqrt{n}E\tilde\xi_1/\sigma_{\xi}\to a$ when $|a|<\infty$;
\item[(iv)]under $H^{(1)}_{a,n}$, if $|a|<+\infty$, then
$n^{-1}\sum_{i=1}^{n}\tilde\xi_i=O_{P}(n^{-1/2})$ and
$n^{-1}\sum_{i=1}^{n}\hat\xi_i=O_{P}(n^{-1/2})$.
\end{itemize}
}
\end{Lemma}



\begin{Lemma}\label{le6}
\rm{
Assume coditions A1 - A3 hold. Under $H^{(1)}_{a,n}$,
 if $|a|<\infty$,
 then
 $\hat\lambda=O_{P}(n^{-1/2}).$
}
\end{Lemma}

\textbf{Proofs of Theorems \ref{th1}-\ref{th1a}}.
Since Theorem~\ref{th1} can be proven along the same line as Theorem~\ref{th1a} (with $a_n=0$),
we omit the proof of Theorem~\ref{th1}.

Case (i): $|a|<\infty$.
 Notice that
$\hat\lambda$ solves the equation
 $\sum_{i=1}^n\hat\xi_i/(1+\hat\lambda\hat\xi_{i})=0$,
which can be rewritten as
$$
0
=\sum_{i=1}^n\hat\xi_i
\{1-\hat\lambda\hat\xi_i+ (\hat\lambda\hat\xi_i)^2/(1+\hat\lambda\hat\xi_i)\}.
$$
Then
\begin{eqnarray}
\hat\lambda&=&\Bigl(\sum_{i=1}\hat\xi_i^2\Bigr)^{-1}\sum_{i=1}^n\hat\xi_i
\{1+(\hat\lambda\hat\xi_i)^2/(1+\hat\lambda\hat\xi_i)\}\nonumber\\
&=&\Bigl(\sum_{i=1}\hat\xi_i^2\Bigr)^{-1}\sum_{i=1}^n\hat\xi_i+
\Bigl(\sum_{i=1}\hat\xi_i^2\Bigr)^{-1}\sum_{i=1}^n\hat\lambda^2\hat\xi_i^3/(1+\hat\lambda\hat\xi_i).\label{th5}
\end{eqnarray}
Applying Taylor's expansion to $\sum_{i=1}^n\log(1+\hat\lambda\hat\xi_i)$ leads to
\begin{eqnarray}
\sum_{i=1}^n\log(1+\hat\lambda\hat\xi_i)
&=&\sum_{i=1}^n \Big\{\hat\lambda\hat\xi_i-(\hat\lambda\hat\xi_i)^2/2+\frac{(\hat\lambda\hat\xi_i)^3}
{3(1+c_{i}\hat\lambda\hat\xi_i)^3}\Big\}\nonumber\\
&=&\hat\lambda\sum_{i=1}^n\hat\xi_i-\hat\lambda\Bigl(\sum_{i=1}^n\hat\xi_i^2\Bigr)\hat\lambda/2
+\hat\lambda^3\sum_{i=1}^n\frac{\hat\xi_i^3}{3(1+c_{i}\hat\lambda\hat\xi_i)^3}, \label{eqk3}
\end{eqnarray}
where $c_{i}\in [0, 1]$.
By Lemma~\ref{le2} and Lemma~\ref{le4}(ii), we obtain that
\begin{eqnarray}
n^{-1}\sum_{i=1}^n\hat\xi_i^2&=&n^{-1}\sum_{i=1}^n\tilde\xi_i^2
+n^{-1}\sum_{i=1}^n(\hat\xi_i^2-\tilde\xi_i^2)
\nonumber\\
&=&E|\tilde\xi_1|^2+o_{P}(1),
\label{Varianceapp}
\end{eqnarray}
which, combined with  $\text{Var}(\tilde\xi_1)=E|\tilde\xi_1|^2-(E\tilde\xi_1)^2$
 and $(E\tilde\xi_1)^2=O(n^{-1})$ in Lemma~\ref{le4}(iii), yields that
 \begin{equation}\label{k2}
 n^{-1}\sum_{i=1}^n\hat\xi_i^2= \text{Var}(\tilde\xi_1) +o_P(1).
 \end{equation}
This, combined with Lemmas~\ref{le4} and~\ref{le6}, implies that
\begin{equation}\label{eqw1}
\Bigl|\hat\lambda^3\sum_{i=1}^n\frac{\hat\xi_i^3}{3(1+c_{i}\hat\lambda\hat\xi_i)^3}\Bigr|
\leq O_{P}(1)|\hat\lambda|^3\max_{1\leq i\leq n}|\hat\xi_i|\sum_{i=1}^n\hat\xi_i^2
=O_{P}(n^{-3/2}n^{1/\gamma}n)=o_{P}(1);
\end{equation}
$$\Bigl|\Bigl(\sum_{i=1}\hat\xi_i^2\Bigr)^{-1}\sum_{i=1}^n\hat\lambda^2\hat\xi_i^3/(1+\hat\lambda\hat\xi_i)\Bigr|\leq
O_{P}(1)\hat\lambda^2\max_{1\leq i\leq n}|\hat\xi_i|=O_{P}(n^{1/\gamma-1}).$$
Let $\bar\xi_n=n^{-1}\sum_{i=1}^{n}\hat\xi_i$. Then,
by Lemmas~\ref{le4}-\ref{le6}, we have
 $|\bar\xi_n|=O_{P}(n^{-1/2})$
 and $|\hat\lambda|=O_{P}(n^{-1/2})$.
Then, it follows from \eqref{th5} and~\eqref{k2} that
$$\hat\lambda=\bar\xi_n/\text{Var}(\tilde\xi_1)+o_{P}(n^{-1/2}).$$
By \eqref{eqk3}, \eqref{k2} and \eqref{eqw1}, we have
$$ 2\sum_{i=1}^n\log(1+\hat\lambda\hat\xi_i)=2n\hat\lambda\bar\xi_n
-n\text{Var}(\tilde\xi_1)\hat\lambda^2+o_{P}(1).$$
Hence,
\begin{eqnarray*}
 R_{n,1}=2\sum_{i=1}^n\log(1+\hat\lambda\hat\xi_i)=n\bar\xi_n^2/\text{Var}(\tilde\xi_1)+o_{P}(1).\nonumber
\end{eqnarray*}
Denoted by $\bar\xi_n^{*}=n^{-1}\sum_{i=1}^n\tilde\xi_i$.
Applying Lemma~\ref{le2}, we obtain that
\begin{eqnarray}
R_{n,1}=2\sum_{i=1}^n\log(1+\hat\lambda\hat\xi_i)=n|\bar\xi_n^{*}|^{2}/\text{Var}(\tilde\xi_1)+o_{P}(1).\label{BAHP}
\end{eqnarray}
Since $\{\tilde\xi_i\}_{i=1}^n$ are iid and $E|\tilde\xi_n|^{\gamma}=O(1)$ for $\gamma>2$ in Lemma~\ref{le4}(ii), by the Lindeberg-Feller central limit theorem,
we establish that $$\sqrt{n}\big(\bar\xi^{*}_{n}-E\tilde\xi_1\big)/\text{Var}(\tilde\xi_1)
=n^{-1/2}\sum_{i=1}^{n}(\tilde\xi_i-E\tilde\xi_1)/\text{Var}(\tilde\xi_1)
\to {\mathcal N}(0,1).$$
Under $H_{a,n}^{(1)}$, we know from  Lemma~\ref{le4}(iii) that
 $\sqrt{n}E\tilde\xi_1/\sqrt{\text{Var}(\tilde\xi_1)}\to a $.
  Therefore,
  $$\sqrt{n}\bar\xi^{*}_{n}/\sqrt{\text{Var}(\tilde\xi_1)}\to
{\mathcal N}(a,1).$$
Then, by \eqref{BAHP},
\begin{equation}\label{ksd1}
 R_{n,1}=2\sum_{i=1}^n\log(1+\hat\lambda\hat\xi_i) \to
\chi_{1}^2(a^2).
\end{equation}

Case (ii): $a=\infty$.
 Let $\hat\lambda_*=n^{-1/2}\text{sgn}(E\tilde\xi_i)$.
By Lemma~\ref{le4}(ii),
$\max_{1\leq i\leq n}\hat\xi_i=o_{P}(n^{1/\gamma})$ for $\gamma>2$.
Then
\begin{equation}\label{eqk1}
\max_{1\leq i\leq n}\hat\lambda_*\hat\xi_i=o_{P}(1).
\end{equation}
 Since
$\hat\lambda=\arg\max_{\lambda}\sum_{i=1}^n\log(1+\lambda\hat\xi_i),$
 we have
$$
 R_{n,1}=2\sum_{i=1}^n\log(1+\hat\lambda\hat\xi_i)
 \geq
2\sum_{i=1}^n\log(1+\hat\lambda_{*}\hat\xi_i).
$$
Then, using \eqref{eqk1} and Taylor's expansion, we establish that
\begin{eqnarray}\label{eqfa1}
R_{n,1}&\ge& 2\sum_{i=1}^n\hat\lambda_{*}\hat\xi_i
-\sum_{i=1}^n\hat\lambda_{*}^2\hat\xi_i^2/(1+c_{i}\hat\lambda_{*}\hat\xi_i)^{2}\nonumber\\
&\geq&2\sum_{i=1}^n\hat\lambda_{*}\hat\xi_i
-2\sum_{i=1}^n\hat\lambda_{*}^2\hat\xi_i^2\{1+o_P(1)\}\nonumber\\
&=&2n^{-1/2}\sum_{i=1}^n\hat\xi_i\text{sgn}(E\tilde\xi_i)
-2n^{-1}\sum_{i=1}^n\hat\xi_i^2\{1+o_P(1)\},
\end{eqnarray}
where
$c_i\in [0,1]$.
By Lemma~\ref{le4}(iii), we have
 $E\tilde\xi_1^2=O(1)$.
 This, combined with $a_{n}\to \infty$ and \eqref{Varianceapp},
 produces that
$n^{-1}\sum_{i=1}^n\hat\xi_i^2=E\tilde\xi_1^2+o_{P}(1)=o_{P}(a_n)$.
Hence,
\begin{equation*}
R_{n,1}\ge 2n^{-1/2}\sum_{i=1}^n\hat\xi_i\text{sgn}(E\tilde\xi_i)+o_P(a_n).
\end{equation*}
Using  Lemma~\ref{le2},
we get
$$
n^{-1}\sum_{i=1}^n\hat\xi_i\text{sgn}(E\tilde\xi_i)=n^{-1}\sum_{i=1}^n\text{sgn}(E\tilde\xi_i)\tilde\xi_i
+o_{P}(a_nn^{-1/2})
=|E\tilde\xi_i|+o_{P}(a_nn^{-1/2}).
$$
Then
\begin{equation}\label{eqka1}
R_{n,1}\ge 2\sqrt{n}|E\tilde\xi_i|+o_P(a_n).
\end{equation}
By Lemma~\ref{le4}(i)-(iii) and condition A3, we know that, for large $n$,
$
|E\tilde\xi_i|> 0.5\sqrt{c_1} |a_n|n^{-1/2}.
$ 
This, together with \eqref{eqka1} and $a_n\to \infty$,
 leads to
\begin{equation}\label{eqfa3}
 P(R_{n,1}\to \infty)\to 1.
 \end{equation}

\textbf{Proofs of Theorems \ref{th2}-\ref{th2a}}.
Since Theorem~\ref{th2}
 can be shown in the same way as Theorem~\ref{th2a}, we only prove Theorem~\ref{th2a}.
The asymptotic results for $R_{n,2}$ follow  along the same line as that for Theorem~\ref{th1a}
by replacing $\hat\xi_i$ and $\tilde\xi_i$ with
$\hat\eta_i$ and $\tilde\eta_i$, respectively.
Since $R_{n,3}=R_{n,2}$, we complete the proof of Theorem~\ref{th2a}.
\hfill$\diamond$

{\bf Proof Lemma~\ref{le1}.}
(i)
We first show that
\begin{eqnarray}
\max_{1\leq i\leq n}\|\bSigma_{n}^{(-i)}-\bSigma_{A}\|_2=O_{P}(\kappa n^{-1/2}).\label{Matrixapp}
\end{eqnarray}
Put $\bSigma_{n}=(n-1)^{-1}\sum_{i=1}^n\bPi(\bX_i)\bPi(\bX_i)^\top$.
By condition A2(iii)
and the inequality
$(E|b|^2)^{1/2}\le (E|b|^{\gamma})^{1/\gamma}$
with $b=\|\bPi(\bX)\|_2^2$
for $\gamma>2$,
we have
  $E\|\bPi(\bX)\|_2^4\leq \{E\|\bPi(\bX)\|_2^{2\gamma}\}^{2/\gamma}=O(\kappa^2)$. Then 
  \begin{eqnarray}
E\Vert \bSigma_{n}-\bSigma_{A}\Vert_{2}^{2}
&\leq& \text{Trace}\{E(\bSigma_{n}-\bSigma_{A})(\bSigma_{n}-\bSigma_{A})\}\nonumber\\
&=&(n-1)^{-2}\sum_{k=1}^{\kappa}\sum_{l=1}^{\kappa}\sum_{i=1}^n[E\{\Pi_{k}^2(\bX_i)\Pi_{l}^2(\bX_i)\}-\Sigma_{A,kl}^2]\nonumber\\
&=& n(n-1)^{-2}\{E\|\bPi(\bX)\|_2^4-\text{Trace}(\bSigma_{A}^2)\}\nonumber\\
&=&O(\kappa^2/n),\label{Mat-Ex}
\end{eqnarray}
where $\Sigma_{A,kl}=E\{\Pi_{k}^2(\bX_i)\Pi_{l}^2(\bX_i)\}.$
Hence,
\begin{equation}\label{eqjl1}
\|\bSigma_{n}-\bSigma_{A}\|_2=O_P(\kappa/\sqrt{n})
\ \ \,\mbox{\rm and}\ \ \,
\|\bSigma_{n}\|_2\le \|\bSigma_{A}\|_2+O_P(\kappa/\sqrt{n}).
\end{equation}
Furthermore,
$$\|\bSigma_{n}^{(-i)}-\bSigma_{A}\|_2\le
\|\bSigma_{n}^{(-i)}-\bSigma_{n}\|_2+\|\bSigma_{n}-\bSigma_{A}\|_2=
(n-1)^{-1}\|\bPi(\bX_i)\|_{2}^2+O_{P}(\kappa/\sqrt{n}).$$
Note that, by  condition A2,
 $E\|\bPi(\bX)\|_2^{2\gamma}=O(\kappa^\gamma)$.
 It follows from Markov's inequality that
\begin{eqnarray}
\max_{1\leq i\leq n}\|\bPi(\bX_i)\|_2^{2\gamma}\leq n n^{-1}\sum_{i=1}^n\|\bPi(\bX_i)\|_2^{2\gamma}=O_{P}(n\kappa^\gamma).\label{maxo}
\end{eqnarray}
That is,
 $\max_{1\leq i\leq n}\|\bPi(\bX_i)\|_2^2=O_{P}(n^{1/\gamma}\kappa)$. Then
 $$\max_{1\leq i\leq n}\|\bSigma_{n}^{(-i)}-\bSigma_{A}\|_2=O_{P}\{(n^{1/\gamma-1}+n^{-1/2})\kappa\}=O_{P}(\kappa n^{-1/2}).$$
That is, \eqref{Matrixapp} holds.
Note that $E\mathrm{y}_{i}\widetilde\bPi(\bX_i)^\top E\{\widetilde\bPi(\bX_1)\mathrm{y}_1\}=\|\bSigma_{A}^{-1/2}E\{\bPi(\bX_{1})\mathrm{y}_{1}\}\|_{2}^2$.
It follows from \eqref{9a} that
\begin{eqnarray*}
E|\tilde\varepsilon_{2,i}|^2=E|\mathrm{y}_{i}-\widetilde\bPi(\bX_i)^\top E\{\widetilde\bPi(\bX_1)\mathrm{y}_1\}|^2
=E|\mathrm{y}_1|^{2}
-\|\bSigma_{A}^{-1/2}E\{\bPi(\bX_{1})\mathrm{y}_{1}\}\|_{2}^2.
\end{eqnarray*}
This, together with condition A2,
 implies that $\|\bSigma_{A}^{-1/2}E\{\bPi(\bX_{1})\mathrm{y}_{1}\}\|_{2}^2\leq E |\mathrm{y}_{1}|^{2}=O(1)$.
 Then, with $\lambda_{\max}(\bSigma_A)=O(1)$ and $\lambda_{\min}(\bSigma_{A})>0$ in condition A2,
 it is easy to see that
\begin{eqnarray}
&& \|E\{\bPi(\bX_{1})\mathrm{y}_{1}\}\|_{2}
=O(1)\,\ \mbox{\rm and}\,\
\|\bSigma_{A}^{-1}E\{\bPi(\bX_{1})\mathrm{y}_{1}\}\|_{2}
=O(1).\label{EPM}
\end{eqnarray}

Denoted by $\bmu_{k}=\bPi(\bX_k)\mathrm{y}_k-E\{\bPi(\bX_k)\mathrm{y}_k\}$.
 Applying Cauchy-Schwarz's inequality,~\eqref{Matrixapp},~\eqref{maxo} and~\eqref{EPM}, we obtain that
\begin{eqnarray*}
|\tilde\varepsilon_{2,i}-\hat\varepsilon_{2,i}|
&\leq&\|\bPi(\bX_i)\|_2
\big\|\frac{1}{n-1}\sum_{k=1(\neq i)}^n(\bSigma_{n}^{(-i)})^{-1}\bmu_{k}
+\{(\bSigma_{n}^{(-i)})^{-1}-\bSigma_{A}^{-1}\}E\{\bPi(\bX_k)\mathrm{y}_k\}\big\|_2\\
&\leq&O_P(n^{\frac{1}{2\gamma}}\sqrt{\kappa})\big\|\frac{1}{n-1}\sum_{k=1(\neq i)}^n\bmu_{k}\big\|_{2}+O_P(n^{\frac{1}{2\gamma}}\sqrt{\kappa})\|(\bSigma_{n}^{(-i)})^{-1}(\bSigma_{n}^{(-i)}-\bSigma_{A})\bSigma_{A}^{-1}\|_{2}\\
&=&O_P(n^{\frac{1}{2\gamma}}\sqrt{\kappa})\big\|\frac{1}{n-1}\sum_{k=1(\neq i)}^n\bmu_{k}\big\|_{2}+O_P(\kappa^{3/2} n^{\frac{1}{2\gamma}-\frac{1}{2}}),
\end{eqnarray*}
uniformly for $1\leq i\leq n$.
Note that
 $E\|\bPi(\bX_1)\|_2^{2\gamma}=O(\kappa^\gamma)$ and
 $E|\mathrm{y}_1|^{2\gamma}=O(1)$ (condition A2), it follows that
\begin{eqnarray}
 E\|\bPi(\bX_1)\mathrm{y}_1\|_{2}^2\leq \sqrt{E\|\bPi(\bX_1)\|_{2}^4E|\mathrm{y}_1|^4}=O(\kappa).\label{TMSC}
\end{eqnarray}
Then
\begin{equation}\label{TMSCa}
 E\big\|\sum_{k=1}^n\bmu_{k}\big\|_2^2
=\sum_{k=1}^nE\|\bmu_{k}\|_{2}^2
=O(n\kappa).
\end{equation}
Recalling that
$E|\mathrm{y}_1|^{2\gamma}=O(1)$, we have
 $\max_{1\leq i\leq n}|\mathrm{y}_i|^{2\gamma}=O_{P}(n)$. This, together with \eqref{maxo}, yields that
\begin{eqnarray}
\max_{1\leq i\leq n}\|\bPi(\bX_i)\mathrm{y}_i\|_{2}\leq \max_{1\leq i\leq n}\|\bPi(\bX_i)\|_{2} \max_{1\leq i\leq n}|\mathrm{y}_i| =O_{P}(n^{1/\gamma}\sqrt{\kappa}).\label{MAXOC}
\end{eqnarray}
Then, applying \eqref{TMSC}-\eqref{MAXOC} and
the inequality
\begin{eqnarray*}
\max_{1\leq i\leq n}\big\|\sum_{k=1(\neq i)}^n\bmu_{k}\big\|_2\leq\big\|\sum_{k=1}^n\bmu_{k}\big\|_2+\max_{1\leq i\leq n}\|\bPi(\bX_i)\mathrm{y}_i\|_{2}+E\|\bPi(\bX_1)\mathrm{y}_1\|_2,
\end{eqnarray*}
we obtain that
\begin{eqnarray}
\max_{1\leq i\leq n}\big\|\frac{1}{n-1}\sum_{k=1(\neq i)}^n\bmu_{k}\big\|_2
=O_{P}(\sqrt{\kappa/n}+n^{1/\gamma-1}\sqrt{\kappa})\label{MAXQ1}.
\end{eqnarray}
Thus, $\max_{1\leq i\leq n}|\tilde\varepsilon_{2,i}-\hat\varepsilon_{2,i}|=O_{P}(\kappa^{3/2}n^{\frac{1}{2\gamma}-\frac{1}{2}})
=O_{P}(n^{\frac{1}{2\gamma}+\frac{3}{4r+2}-\frac{1}{2}})$.
Similarly, we can also show that $\max_{1\leq i\leq n}|\tilde\varepsilon_{1,i}-\hat\varepsilon_{1,i}|
=O_{P}(n^{\frac{1}{2\gamma}+\frac{3}{4r+2}-\frac{1}{2}})$.

(ii)
 Let $\hat\varepsilon_{2,i}^*=\frac{1}{n-1}\sum_{k=1}^n\bPi(\bX_i)^\top\bSigma_{n}^{-1}\bPi(\bX_k)\mathrm{y}_k$.
Then
$$n^{-1}\sum_{i=1}^{n}|\hat\varepsilon_{2,i}-\tilde\varepsilon_{2,i}|^2\leq 2n^{-1}\sum_{i=1}^{n}|\hat\varepsilon_{2,i}^*-\tilde\varepsilon_{2,i}|^2+
2n^{-1}\sum_{i=1}^{n}|\hat\varepsilon_{2,i}-\hat\varepsilon_{2,i}^*|^2.$$
By the inequality $(a+b)^2\leq 2(a^2+b^2)$, we have
\begin{eqnarray*}
n^{-1}\sum_{i=1}^{n}|\hat\varepsilon_{2,i}-\hat\varepsilon_{2,i}^*|^2
&\leq& 2n^{-1}\sum_{i=1}^{n}\bigl|
\frac{1}{n-1}\sum_{k=1(\neq i)}^n\bPi(\bX_i)^\top\bigl[\{\bSigma_{n}^{(-i)}\}^{-1}-\bSigma_{n}^{-1}\bigr]\bPi(\bX_k)\mathrm{y}_k\bigr|^2\\
&&+\frac{2}{n(n-1)^2}\sum_{i=1}^{n}\bigl|\bPi(\bX_i)^\top\bSigma_{n}^{-1}\bPi(\bX_i)\mathrm{y}_i\bigr|^2\\
&\equiv&r_{n,1}+r_{n,2}.
\end{eqnarray*}
Applying~\eqref{EPM} and ~\eqref{MAXQ1}, we obtain that
 $$\max_{1\leq i\leq n}\bigl\|\frac{1}{n-1}\sum_{k=1(\neq i)}^n\bmu_k+E\bPi(\bX_1)\mathrm{y}_1\bigr\|_{2}=O_{P}(1).$$
This, combined with the Cauchy-Schwarz inequality,~\eqref{Matrixapp} and~\eqref{eqjl1}, yields that
\begin{eqnarray*}
r_{n,1}&\leq& 2n^{-1}\sum_{i=1}^{n}\|\bPi(\bX_i)\|_{2}^2\cdot\|\{\bSigma_{n}^{(-i)}\}^{-1}-\bSigma_{n}^{-1}\|_{2}^{2}\cdot
\bigl\|\frac{1}{n-1}\sum_{k=1(\neq i)}^n\bmu_k+E\bPi(\bX_1)\mathrm{y}_1\bigr\|_{2}^2\\
&\leq& O_{P}(1)\frac{1}{(n-1)^2n}\sum_{i=1}^{n}\|\bPi(\bX_i)\|_{2}^2\cdot\|\bPi(\bX_i)\|_{2}^4\cdot\|\bSigma_{n}^{-1}\|_{2}^2\cdot
\|\{\bSigma_{n}^{(-i)}\}^{-1}\|_{2}^2\\
&=&O_{P}(n^{-2})n^{-1}\sum_{i=1}^{n}\|\bPi(\bX_i)\|_{2}^6
\end{eqnarray*}
and
\begin{eqnarray*}
r_{n,2}&\leq& \frac{2}{n(n-1)^2}\sum_{i=1}^{n}\|\bPi(\bX_i)\|_{2}^2\cdot \|\bSigma_{n}^{-1}\|_{2}^2\cdot\|\bPi(\bX_i)\mathrm{y}_i\|_2^2=
O_{P}(n^{-2})n^{-1}\sum_{i=1}^{n}\|\bPi(\bX_i)\|_{2}^4\mathrm{y}_i^2.
\end{eqnarray*}
By condition A2, we have
 $\max_{1\leq i\leq n}|\mathrm{y}_i|=O_{P}(n^{1/(2\gamma)})$
 and
 $$E\|\bPi(\bX_{1})\|_{2}^4\leq \{E\|\bPi(\bX_{1})\|_{2}^{2\gamma}\}^{2/\gamma}=O(\kappa^2).$$
 Then, applying~\eqref{maxo},
  we establish that
 $$n^{-1}\sum_{i=1}^{n}\|\bPi(\bX_i)\|_{2}^6
\leq \max_{1\leq i\leq n}\|\bPi(\bX_i)\|_{2}^2 n^{-1}\sum_{i=1}^{n}\|\bPi(\bX_i)\|_{2}^4
=O_{P}(\kappa^3 n^{1/\gamma})$$ and
$$n^{-1}\sum_{i=1}^{n}\|\bPi(\bX_i)\|_{2}^4\mathrm{y}_i^2\leq \max_{1\leq i\leq n}\mathrm{y}_i^2 n^{-1}\sum_{i=1}^{n}\|\bPi(\bX_i)\|_{2}^4=O_{P}(\kappa^2n^{1/\gamma}).$$
Thus,
 $n^{-1}\sum_{i=1}^{n}|\hat\varepsilon_{2,i}-\hat\varepsilon_{2,i}^*|^2
=O_{P}(\kappa^3 n^{1/\gamma-2})$.
It follows that
\begin{equation}\label{eqq1}
 n^{-1}\sum_{i=1}^{n}|\hat\varepsilon_{2,i}-\tilde\varepsilon_{2,i}|^2\leq 2n^{-1}\sum_{i=1}^{n}|\hat\varepsilon_{2,i}^*-\tilde\varepsilon_{2,i}|^2+
O_{P}(\kappa^3 n^{1/\gamma-2}).
\end{equation}
Put
$\bv=\bSigma_{A}^{-1}E\{\bPi(\bX_k)\mathrm{y}_k\}$. Then
\begin{eqnarray}
\hat\varepsilon_{2,i}^*-\tilde\varepsilon_{2,i}
&=&\frac{1}{n-1}\sum_{k=1}^n\bigl[\bPi(\bX_i)^\top\bSigma_{n}^{-1}\bmu_{k}
+\bPi(\bX_i)^\top\{\bSigma_{n}^{-1}-\bSigma_{A}^{-1}\}E\{\bPi(\bX_k)\mathrm{y}_k\}\bigr]\nonumber\\
&=&\frac{1}{n-1}\sum_{k=1}^n\bPi(\bX_i)^\top\bSigma_{n}^{-1}\bmu_{k}
+\frac{1}{n-1}\sum_{k=1}^n\bPi(\bX_i)^\top\bSigma_{n}^{-1}(\bSigma_{A}-\bSigma_{n})\bv\nonumber\\
&\equiv&I_{i,1}+I_{i,2}.  \label{eqp0}
\end{eqnarray}
By the Cauchy-Schwzarz inequality,~\eqref{eqjl1}, ~\eqref{EPM},~\eqref{TMSCa}, $\lambda_{\max}(\bSigma_{A})=O(1)$,
and
$\lambda_{\min}(\bSigma_A)>0$
in condition A2, it holds that
\begin{equation}\label{eqp1}
n^{-1}\sum_{i=1}^nI_{i,2}^2\leq \frac{(n-1)}{n} \|\bSigma_{n}\|_{2}\cdot\|\bSigma_{n}^{-1}\|_{2}^2\cdot
\|(\bSigma_{A}-\bSigma_{n})\|_{2}^2\cdot\|\bv\|_{2}^2=O_{P}(\kappa^2/n);
\end{equation}
\begin{equation}
n^{-1}\sum_{i=1}^nI_{i,1}^2
\leq \frac{(n-1)}{n}\|\bSigma_{n}\|_{2}\cdot \|\bSigma_{n}^{-1}\|_{2}^2\cdot
\frac{1}{(n-1)^2}\|\sum_{k=1}^n\bmu_{k}\|_{2}^2
=O_{P}(\kappa/n). \label{eqp2}
\end{equation}
Naturally, combining \eqref{eqp0}-\eqref{eqp2}
leads to
$$n^{-1}\sum_{i=1}^{n}|\hat\varepsilon_{2,i}^*-\tilde\varepsilon_{2,i}|^2
\le 2 n^{-1}\sum_{i=1}^n I_{i,1}^2 + 2 n^{-1} \sum_{i=1}^n I_{i,2}^2
=O_{P}(\kappa^2/n),$$
which, together with \eqref{eqq1}, yields that
 $n^{-1}\sum_{i=1}^{n}|\hat\varepsilon_{2,i}-\tilde\varepsilon_{2,i}|^2
 =O_P(n^{2/(2r+1)-1}).
 $
Along the same line, we can also show that $n^{-1}\sum_{i=1}^{n}|\hat\varepsilon_{1,i}-\tilde\varepsilon_{1,i}|^2
 =O_P(n^{2/(2r+1)-1}).
 $
\hfill$\diamond$\\


\textbf{Proof of Lemma~\ref{le2}}.
(i)
By the definitions of $\hat\xi_i$ and $\tilde\xi_i$,
we have
$$n^{-1}\sum_{i=1}^n(\hat\xi_i-\tilde\xi_i)
=
n^{-1}\sum_{i=1}^n(\hat\varepsilon_{1,i}^2-\tilde\varepsilon_{1,i}^2)
-n^{-1}\sum_{i=1}^n(\hat\varepsilon_{2,i}^2-\tilde\varepsilon_{2,i}^2).$$
We will show each term on the righthand side of the above equation
is $o_P(n^{-1/2}).$
In the following we only show this for the 2nd term, since it can be done similarly for the 1st term.

Notice that
\begin{eqnarray*}
|\hat\varepsilon_{2,i}|^2-|\tilde\varepsilon_{2,i}|^2
=(\hat\varepsilon_{2,i}-\tilde\varepsilon_{2,i})^2+2\tilde\varepsilon_{2,i}(\hat\varepsilon_{2,i}-\tilde\varepsilon_{2,i}).
 \label{DIFFE}
\end{eqnarray*}
By Lemma~\ref{le1},
 $n^{-1}\sum_{i=1}^n|\hat\varepsilon_{2,i}-\tilde\varepsilon_{2,i}|^2=o_{P}(n^{-1/2})$.
Then it suffices to show that
\begin{eqnarray}
n^{-1}\sum_{i=1}^n\tilde\varepsilon_{2,i}(\tilde\varepsilon_{2,i}-\hat\varepsilon_{2,i})=o_{P}(n^{-1/2}). \label{copr}
\end{eqnarray}

The reader who does not wish to study the lengthy proof may skip to the proof of (ii).
Let
 $\bdelta_{i}=\widetilde\bPi(\bX_i)[
 \mathrm{y}_i-\widetilde\bPi(\bX_i)^\top E\{\widetilde\bPi(\bX_1)\mathrm{y}_1\}]
 =\tilde\varepsilon_{2,i}\widetilde\bPi(\bX_i).$
Then
 $E\bdelta_{i}^\top\bdelta_{j}=0$ for $i\neq j$,
and
\begin{equation}\label{eqx1}
E\|\bdelta_{i}\|_{2}^2\leq 2E\|\widetilde\bPi(\bX_i)\mathrm{y}_i\|_{2}^2+2\|\bSigma_{A}^{-1}E\{\bPi(\bX_1)\mathrm{y}_1\}\|_{2}^2\cdot
E\{\|\bPi(\bX_i)\|_{2}^2\cdot\|\widetilde\bPi(\bX_i)\|_{2}^2\}.
\end{equation}
Note that
$E\|\bPi(\bX_i)\|_{2}^4\leq \{E\|\bPi(\bX_i)\|_{2}^{2\gamma}\}^{2/\gamma}=O(\kappa^2)$
and $\lambda_{\min}(\bSigma_{A})>0$.
It follows from
 \eqref{EPM},~\eqref{TMSC} and \eqref{eqx1}
 that
$E\|\bdelta_{i}\|_{2}^2=O(\kappa^2)$.
Hence,
 \begin{equation}\label{L2Vaa}
 \big\|n^{-1}\sum_{i=1}^n\tilde\varepsilon_{2,i}\widetilde\bPi(\bX_i)\big\|_{2}=O_{P}(\kappa n^{-1/2}).
 \end{equation}
 Using the Cauchy-Schwarz inequality,~\eqref{EPM},
 $\lambda_{\min}(\bSigma_{A})>0$ and $E\|\bPi(\bX_i)\|_{2}^{2\gamma}=O(\kappa^{\gamma})$ for $\gamma>2$, we establish that
\begin{eqnarray*}
E|\tilde\varepsilon_{2,i}|\cdot\|\widetilde\bPi(\bX_i)\|_{2}^2&\leq&E\|\widetilde\bPi(\bX_i)\|_{2}^2|\mathrm{y}_i|
+\|\bSigma_{A}^{-1}E\{\bPi(\bX_1)\mathrm{y}_1\}\|_{2}\cdot E\{\|\widetilde\bPi(\bX_i)\|_{2}^2\cdot\|\bPi(\bX_i)\|_{2}\}\\
&\leq&O(1)\sqrt{E\|\bPi(\bX_i)\|_{2}^4E\mathrm{y}_i^2}+O(1)E\|\bPi(\bX_i)\|_{2}^3\\
&=&O(\kappa^{3/2}).
\end{eqnarray*}
Thus,
\begin{eqnarray}
  n^{-1}\sum_{i=1}^n|\tilde\varepsilon_{2,i}|\cdot\|\widetilde\bPi(\bX_i)\|_{2}^2=O_{P}(\kappa^{3/2}).\label{L2V}
\end{eqnarray}
Denoted by $\mF_{i}=\bSigma_{A}^{1/2}\{(\bSigma_{n}^{(-i)})^{-1}-\bSigma_{A}^{-1}\}\bSigma_{A}^{1/2},$
$\mF=\bSigma_{A}^{1/2}\{\bSigma_{n}^{-1}-\bSigma_{A}^{-1}\}\bSigma_{A}^{1/2}$, $\bG_{i}=(n-1)^{-1}\sum_{k=1(\neq i)}^n\widetilde\bPi(\bX_k)\mathrm{y}_k,$
 $\bG=(n-1)^{-1}\sum_{k=1}^n\widetilde\bPi(\bX_k)\mathrm{y}_k$,
 $\bL_i=\frac{1}{n-1}\sum_{k=1(\neq i)}^n
 [\widetilde\bPi(\bX_k)\mathrm{y}_k
 -E\{\widetilde\bPi(\bX_k)\mathrm{y}_k\}],$
  and
$\bL=\frac{1}{n-1}\sum_{k=1}^n[\widetilde\bPi(\bX_k)\mathrm{y}_k-E\{\widetilde\bPi(\bX_k)\mathrm{y}_k\}]$.
Let

  $ I_1=n^{-1}\sum_{i=1}^n\tilde\varepsilon_{2,i}\widetilde\bPi(\bX_i)^\top\mF\bG,$

$I_2=n^{-1}\sum_{i=1}^n\tilde\varepsilon_{2,i}\widetilde\bPi(\bX_i)^\top\{-(\mF_i-\mF)(\bG_i-\bG)\},$

$I_3=n^{-1}\sum_{i=1}^n\tilde\varepsilon_{2,i}\widetilde\bPi(\bX_i)^\top\mF_i(\bG_i-\bG),$

$I_4=n^{-1}\sum_{i=1}^n\tilde\varepsilon_{2,i}\widetilde\bPi(\bX_i)^\top(\mF_i-\mF)\bG_i,$

$I_5=n^{-1}\sum_{i=1}^n\tilde\varepsilon_{2,i}\widetilde\bPi(\bX_i)^\top(\bL_i-\bL),$
and
$I_6=n^{-1}\sum_{i=1}^n\tilde\varepsilon_{2,i}\widetilde\bPi(\bX_i)^\top\bL.$

\noindent Then it can be rewritten that
\begin{eqnarray*}
n^{-1}\sum_{i=1}^n\tilde\varepsilon_{2,i}(\hat\varepsilon_{2,i}-\tilde\varepsilon_{2,i})
 &=&n^{-1}\sum_{i=1}^n\tilde\varepsilon_{2,i}\widetilde\bPi(\bX_i)^\top\mF_{i}\bG_{i}
 +n^{-1}\sum_{i=1}^n\tilde\varepsilon_{2,i}\widetilde\bPi(\bX_i)^\top\bL_i
 = \sum_{j=1}^6 I_j.
 \end{eqnarray*}
Hence,
by \eqref{copr},
 it suffices to show that $I_j=o_P(n^{-1/2})$.
 Applying the Cauchy-Schwarz inequality and \eqref{L2Vaa}-\eqref{L2V}, we establish that
$$
 |I_{1}|\leq \Big\|n^{-1}\sum_{i=1}^n\tilde\varepsilon_{2,i}\widetilde\bPi(\bX_i)\Big\|_{2}\|\mF\|_{2}\|\bG\|_{2}
 =O_{P}(\kappa /\sqrt{n})\|\mF\|_{2}\|\bG\|_{2};
$$
\begin{eqnarray*}
 |I_3|&\leq& \frac{1}{n(n-1)}\sum_{i=1}^n\big\|\tilde\varepsilon_{2,i}\widetilde\bPi(\bX_i)\big\|_{2}\cdot
 \|\mF_i\|_{2}\cdot\|\widetilde\bPi(\bX_i)\mathrm{y}_i\|_2\\
 &\leq&\max_{1\leq i\leq n} \|\mF_i\|_{2}|\mathrm{y}_i|
  \frac{1}{n(n-1)}\sum_{i=1}^n|\tilde\varepsilon_{2,i}|\cdot\big\|\widetilde\bPi(\bX_i)\big\|_{2}^2
 \\
 &=&O_{P}(\kappa^{3/2}/n) \max_{1\leq i\leq n}|\mathrm{y}_i| \max_{1\le i\le n}\|\mF_i\|_{2}.
\end{eqnarray*}
Combining~\eqref{Matrixapp},~\eqref{eqjl1}, $E|\mathrm{y}_i|^{\gamma}=O(1)$, $\lambda_{\min}(\bSigma_{A})>0$ and $\lambda_{\max}(\bSigma_{A})=O(1)$, we arrive at $\max_{1\leq i\leq n}|\mathrm{y}_i|=O_{P}(n^{\frac{1}{2\gamma}})$,
 $$\max_{1\leq i\leq n}\|\mF_i\|_{2}\leq \max_{1\leq i\leq n}\|\bSigma_{A}^{1/2}\|_{2}^{2}\cdot\|\bSigma^{(-i)}_{n}\|_{2}\cdot\|\bSigma_{A}^{-1}\|_{2}\cdot
\|\bSigma^{(-i)}_{n}-\bSigma_{A}\|_{2}=O_{P}(\kappa/\sqrt{n}),$$ and
$\|\mF\|_{2}\leq\|\bSigma_{A}^{1/2}\|_{2}^{2}\cdot\|\bSigma_{n}\|_{2}\cdot\|\bSigma_{A}^{-1}\|_{2}\cdot
\|\bSigma_{n}-\bSigma_{A}\|_{2}=O_{P}(\kappa/\sqrt{n}).$
 Using \eqref{EPM} and  \eqref{TMSC},
 we
get
\begin{eqnarray*}
E\|\bG\|_{2}^2&=& \frac{1}{(n-1)^2}\sum_{i=1}^n\sum_{j=1}^n E\widetilde\bPi(\bX_i)^\top\widetilde\bPi(\bX_j)\mathrm{y}_i\mathrm{y}_j\\
&=&\frac{n}{(n-1)^2}E\|\widetilde\bPi(\bX_j)\mathrm{y}_i\|_{2}^2+\frac{n}{(n-1)}
\|E\widetilde\bPi(\bX_j)\mathrm{y}_i\|_{2}^2\\
&=&O(1).
\end{eqnarray*}
Thus,
 $\|\bG\|_{2}=O_P(1)$.
 Since $\gamma>2$ and $r>1.5$, we have
$I_{1}=O_{P}(\kappa^2/n)
=o_{P}(n^{-1/2})$ and
$I_{3}
=O_{P}(n^{\frac{1}{2\gamma}}\kappa^{5/2}n^{-3/2})
=o_{P}(n^{-1/2})$.
Similarly, we can show that $I_j=o_P(n^{-1/2})$ for $j=2,4,5,6$.

(ii)  Notice that
$$\hat\xi_i^2-\tilde\xi_i^2
=(\hat\xi_i-\tilde\xi_i)^2+2\tilde\xi_i(\hat\xi_i-\tilde\xi_i).
$$
It follows from the Cauchy-Schwarz inequality that
\begin{eqnarray*}
 \bigl|n^{-1}\sum_{i=1}^n( \hat\xi_i^2-\tilde\xi_i^2)\bigr|
&\leq& n^{-1}\sum_{i=1}^n(\hat\xi_i-\tilde\xi_i)^2
+2|\tilde\xi_i|\cdot|\hat\xi_i-\tilde\xi_i|\\
&\leq&
2n^{-1}\Bigl(\sum_{i=1}^n|\tilde\xi_i|^2\sum_{i=1}^n|\hat\xi_i-\tilde\xi_i|^2\Bigr)^{1/2}
+n^{-1}\sum_{i=1}^n(\hat\xi_i-\tilde\xi_i)^2.
\end{eqnarray*}
By Jensen's inequality and
 condition A3, we obtain that
\begin{eqnarray}
E|\tilde\xi_1|^\gamma=E|\tilde\varepsilon_{1,1}^2-\tilde\varepsilon^2_{2,1}|^\gamma\leq 2^{\gamma-1}E|\tilde\varepsilon_{1,1}|^{2\gamma}+2^{\gamma-1}E|\tilde\varepsilon_{2,1}|^{2\gamma}=O(1).\label{MSAPE}
\end{eqnarray}
Thus, by condition A3 and Markov's inequality,
 $n^{-1}\sum_{i=1}^n\tilde\xi_i^2=O_{P}(1)$.
To complete the proof, it's sufficient to show that
 $n^{-1}\sum_{i=1}^n |\hat\xi_i-\tilde\xi_i|^2=o_{P}(1)$.

By the definitions of $\hat\xi_i$, $\tilde{\xi}_i$, and Jessen's inequality, we establish that
\begin{eqnarray*}
n^{-1}\sum_{i=1}^n(\hat\xi_i-\tilde\xi_i)^2
&=& n^{-1}\sum_{i=1}^n \bigl\{(\tilde\varepsilon_{1,i}-\hat\varepsilon_{1,i})^2-(\tilde\varepsilon_{2,i}-\hat\varepsilon_{2,i})^2\\
&&
+2\tilde\varepsilon_{1,i}(\hat\varepsilon_{1,i}-\tilde\varepsilon_{1,i})
-2\tilde\varepsilon_{2,i}(\hat\varepsilon_{2,i}-\tilde\varepsilon_{2,i})\bigr\}^2\\
&\leq&\frac{4}{n}\sum_{i=1}^n\sum_{j=1}^2(\tilde\varepsilon_{j,i}-\hat\varepsilon_{j,i})^4
+\frac{16}{n}\sum_{i=1}^n\sum_{j=1}^2 \tilde\varepsilon_{j,i}^2(\tilde\varepsilon_{j,i}-\hat\varepsilon_{j,i})^2.
\end{eqnarray*}
Applying Lemma~\ref{le1}, we establish that
\begin{eqnarray*}
n^{-1}\sum_{i=1}^n\sum_{j=1}^2(\tilde\varepsilon_{j,i}-\hat\varepsilon_{j,i})^4
&\leq& \max_{1\leq i\leq n,1\leq j\leq 2}(\tilde\varepsilon_{j,i}-\hat\varepsilon_{j,i})^2
n^{-1}\sum_{i=1}^n\sum_{j=1}^2(\tilde\varepsilon_{j,i}-\hat\varepsilon_{j,i})^2\\
&=&O_{P}(n^{\frac{1}{\gamma}+\frac{5}{2r+1}-2});
\end{eqnarray*}
\begin{eqnarray*}
n^{-1}\sum_{i=1}^n\sum_{j=1}^2 \tilde\varepsilon_{j,i}^2(\tilde\varepsilon_{j,i}-\hat\varepsilon_{j,i})^2
&\leq& \max_{1\leq i\leq n, 1\leq j\leq 2}\tilde\varepsilon_{j,i}^2
n^{-1}\sum_{i=1}^n\sum_{j=1}^2(\tilde\varepsilon_{j,i}-\hat\varepsilon_{j,i})^2\\
&=&\max_{1\leq i\leq n, 1\leq j\leq 2}\tilde\varepsilon_{j,i}^2 \, O_{P}(n^{\frac{2}{2r+1}-1}).
\end{eqnarray*}
By condition A3,
$E|\tilde\varepsilon_{2,1}|^{2\gamma}=O(1)$
 for $\gamma>2$. Then, by the Markov inequality,
$$P(n^{-1}\max_{1\leq i\leq n}|\tilde\varepsilon_{2,i}|^{2\gamma}>M_n)
\le M_n^{-1}n^{-1}\sum_{i=1}^n E|\tilde\varepsilon_{2,i}|^{2\gamma}\to 0,$$
as $M_n\to \infty$.
 Thus,
$\max_{1\leq i\leq n}|\tilde\varepsilon_{2,i}|^2=O_{P}(n^{1/\gamma})$.
Similarly, $\max_{1\leq i\leq n}|\tilde\varepsilon_{1,i}|^2=O_{P}(n^{1/\gamma})$.
Hence,
\begin{eqnarray}
\max_{1\leq i\leq n, 1\leq j\leq 2}|\tilde\varepsilon_{j,i}|^2=O_{P}(n^{1/\gamma}).\label{MAXO}
\end{eqnarray}
Combining the above results with $\gamma>2$ and $r>1.5$ leads to
$$
n^{-1}\sum_{i=1}^n(\hat\xi_i-\tilde\xi_i)^2
= O_{P}(n^{\frac{1}{\gamma}+\frac{5}{2r+1}-2})+O_{P}(n^{\frac{1}{\gamma}+\frac{2}{2r+1}-1})
=o_{P}(1).
$$
\hfill$\diamond$\\


\textbf{Proof of Lemma~\ref{le4}}.
(i) By Lemma~\ref{le2}, we have
\begin{equation}\label{eqsat2}
n^{-1}\sum_{i=1}^n(\tilde\xi_i-\hat\xi_i)=o_{P}(n^{-1/2})\,\,\
\mbox{\rm and}\,\,\
n^{-1}\sum_{i=1}^n(\hat\xi_i^2-\tilde\xi_i^2)=o_{P}(1).
\end{equation}
Let $\bomega_{1,n}=n^{-1}\sum_{i=1}^n(\hat\xi_i^2-\tilde\xi_i^2)/(E\hat\xi_i^2+E\tilde\xi_i^2)$.
By condition A3, we have
 $E\tilde\xi_i^2\geq \text{Var}(\tilde\xi_i)>c_1$.
 This, combined with~\eqref{eqsat2}, ensures that
 $$\bomega_{1,n}=o_{P}(1)\,\,\
 \mbox{\rm and }\,\,\
 \sup_{n}E|\bomega_{1,n}|\leq 1.
 $$
 Applying Theorem A (Serfling, 1980, page 14),
 we obtain that $|E\bomega_{1,n}|\leq E|\bomega_{1,n}|\to 0$.
 Then, it is easy to see that
\begin{eqnarray}\label{eqsat4}
 \ E\hat\xi_i^2/E\tilde\xi_i^2\to 1.
\end{eqnarray}
Let
$\mathcal{X}_n=n^{-1}\sum_{i=1}^n (\hat\xi_i-E\tilde\xi_i)$ and
$\mathcal{Y}_n=n^{-1}\sum_{i=1}^n (\tilde\xi_i-E\tilde\xi_i)$.
Using the identity
 $\mathcal{X}_n^2-\mathcal{Y}_n^2=(\mathcal{X}_n-\mathcal{Y}_n)^2+2\mathcal{Y}_n(\mathcal{X}_n-\mathcal{Y}_n)$, we get
\begin{eqnarray}
\mathcal{X}_n^2-\mathcal{Y}_n^2
=\bigl\{n^{-1}\sum_{i=1}^n (\hat\xi_i-\tilde\xi_i)\bigr\}^2
+2\mathcal{Y}_n
n^{-1}\sum_{i=1}^n (\hat\xi_i-\tilde\xi_i\bigr). \label{eqll1}
\end{eqnarray}
By Markov's inequality and \eqref{MSAPE}, it holds that
$$P\Big(|\mathcal{Y}_n|>c_n n^{-1/2}\Big)\leq n^{-1}c_n^{-2}\sum_{i=1}^nE(\tilde\xi_i-E\tilde\xi_i)^2
\to 0 $$
for any $c_n\to\infty$.
Hence,
\begin{eqnarray}\label{eqsat6}
\mathcal{Y}_n
=O_{P}(n^{-1/2}).
\end{eqnarray}
This, combined with~\eqref{eqsat2} and \eqref{eqll1}, yields that
$$\mathcal{X}_n^2-\mathcal{Y}_n^2
=o_{P}(n^{-1}).$$
Define $\bomega_{2,n}=(\mathcal{X}_n^2-\mathcal{Y}_n^2)/(E\mathcal{X}_n^2+E\mathcal{Y}_n^2).$
 Since $E\mathcal{Y}_n^2= n^{-1}\text{Var}(\tilde\xi_1)\geq n^{-1}c_1$,
 we have
  $\bomega_{2,n}=o_{P}(1)$ and $\sup_nE|\bomega_{2,n}|\leq 1$.
  Hence,
 $E \bomega_{2,n}=o(1)$.
 Similar to~\eqref{eqsat4}, we get
\begin{eqnarray}\label{eqsat7}
 E\mathcal{X}_n^2/E\mathcal{Y}_n^2
  \to 1.
\end{eqnarray}
Denoted by $\bomega_{3,n}=\{n\text{Var}(\tilde\xi_1)\}^{-1/2}\sum_{i=1}^n(\tilde\xi_i-\hat\xi_i)$.
Then, by~\eqref{eqsat2} and $\text{Var}(\tilde\xi_1)>c_1$,
 $\bomega_{3,n}=o_{P}(1)$.
 Furthermore,
\begin{eqnarray*}
E|\bomega_{3,n}|^2&=&\{\text{Var}(\tilde\xi_1)/n\}^{-1}E(\mathcal{X}_n-\mathcal{Y}_n)^2\\
&\leq&2(E\mathcal{Y}_n^2)^{-1}(E\mathcal{X}_n^2+E\mathcal{Y}_n^2)\\
&=&2+2E\mathcal{X}_n^2/E\mathcal{Y}_n^2.
\end{eqnarray*}
It follows from \eqref{eqsat7} that
 $\sup_nE|\bomega_{3,n}|^2$ is bounded.
Thus,
 \begin{eqnarray}\label{eqsat5}
E\bomega_{3,n}
\to 0,\,\, \mbox{\rm or equivalently} \,\,
\sqrt{n}(E\tilde\xi_1  - E\hat\xi_1)/\sqrt{\text{Var}(\tilde\xi_1)}\to 0.
 \end{eqnarray}
Since $\text{Var}(\tilde\xi_1)>c_1>0$,
$E(\tilde\xi_i-\hat\xi_i) \to 0.$
This, combined with~\eqref{eqsat4}, yields that
 $\text{\Var}(\hat\xi_i)- \text{\Var}(\tilde\xi_i)=o(1)$,
 or equivalently
 \begin{equation}\label{eqsat1}
 \text{\Var}(\tilde\xi_i)=\sigma_{\xi}^2+o(1).
 \end{equation}

(ii)
 By~\eqref{MSAPE},  we obtain that
$
\max_{1\leq i\leq n}|\tilde\xi_i|^\gamma\leq \sum_{i=1}^n|\tilde\xi_i|^\gamma=O_{P}(n),
$
which implies that
$\max_{1\leq i\leq n}|\tilde\xi_i|=O_{P}(n^{1/\gamma})$.
Using
Lemma~\ref{le1}(i) and the definitions of $\hat\xi_i$ and $\tilde\xi_i$, we obtain that
\begin{eqnarray*}
|\hat\xi_i-\tilde\xi_i|
&=& \bigl|
(\tilde\varepsilon_{1,i}-\hat\varepsilon_{1,i})^2-(\tilde\varepsilon_{2,i}-\hat\varepsilon_{2,i})^2
+2\tilde\varepsilon_{1,i}(\hat\varepsilon_{1,i}-\tilde\varepsilon_{1,i})
-2\tilde\varepsilon_{2,i}(\hat\varepsilon_{2,i}-\tilde\varepsilon_{2,i})
\bigr|\\
&\leq&2\max_{1\leq i\leq n, 1\leq j\leq 2}(\tilde\varepsilon_{j,i}-\hat\varepsilon_{j,i})^2
+4\max_{1\leq i\leq n, 1\leq j\leq 2}|\tilde\varepsilon_{j,i}(\hat\varepsilon_{j,i}-\tilde\varepsilon_{j,i})|\\
&=&O_{P}(n^{\frac{1}{\gamma}+\frac{3}{2r+1}-1})+
O_{P}(n^{\frac{1}{2\gamma}+\frac{3}{4r+2}-\frac{1}{2}})\max_{1\leq i\leq n, 1\leq j\leq 2}|\tilde\varepsilon_{j,i}|,
\end{eqnarray*}
which, combined with
\eqref{MAXO}
and $r>1.5$, 
yields that
$\max_{1\leq i\leq n}|\hat\xi_i-\tilde\xi_i|=O_{P}(n^{1/\gamma})$.
Thus, $\max_{1\leq i\leq n}|\hat\xi_{i}|\leq \max_{1\leq i\leq n}|\tilde\xi_{i}|+\max_{1\leq i\leq n}|\hat\xi_i-\tilde\xi_i|=O_{P}(n^{1/\gamma})$.
Note that $\tilde\xi_i$ are iid. It follows from \eqref{MSAPE} that $n^{-1}\sum_{i=1}^n\tilde\xi_i^2=E\tilde\xi_i^2+o_{P}(1)$.

(iii).
Case 1: $|a|=+\infty$.
By~\eqref{eqsat5} and~\eqref{eqsat1}, we have
$$a_n^{-1}\sqrt{n}E(\hat\xi_i/\sigma_{\xi}-\tilde\xi_i/\sigma_{\xi})\to 0.$$
 Under $H^{(1)}_{a,n},$
we know that
 $a_n^{-1}\sqrt{n}E\hat\xi_i/\sigma_{\xi}=1$
 and
   $a_n^{-1}\sqrt{n}E\tilde\xi_i/\sigma_{\xi}\to 1$.

Case 2: $|a|<+\infty$.
By~\eqref{eqsat5} and~\eqref{eqsat1}, we have
 $\sqrt{n}E(\hat\xi_i/\sigma_{\xi}-\tilde\xi_i/\sigma_{\xi})\to 0$. Thus,
 under $H^{(1)}_{a,n},$
we get
 \begin{equation}\label{eqpp1}
 \sqrt{n}E\hat\xi_i/\sigma_{\xi}=a_n\to a\,\,\
 \mbox{\rm and}\,\,\
   \sqrt{n}E\tilde\xi_i/\sigma_{\xi}\to a.
   \end{equation}

(iv). By \eqref{eqpp1},
we have $E(\tilde\xi_1)/\sigma_{\xi}=O(n^{-1/2})$ when $|a|<\infty$. Then
\begin{eqnarray}
n^{-1}\sum_{i=1}^{n}\tilde\xi_i/\sigma_{\xi}
&=&E(\tilde\xi_1/\sigma_{\xi})+
\frac{1}{n\sigma_{\xi}}\sum_{i=1}^{n}(\tilde\xi_i-E\tilde\xi_i)\nonumber\\
&=& \frac{1}{n\sigma_{\xi}}\sum_{i=1}^{n}(\tilde\xi_i-E\tilde\xi_i)
+O(n^{-1/2}).
\end{eqnarray}
By condition A3,  $\text{Var}(\tilde\xi_{1})>c_1$.
It follows from \eqref{eqsat1} that
 $\sigma_{\xi}\geq c_{1}+o(1)$.
This, combined with~\eqref{eqsat6}, yields that $n^{-1}\sum_{i=1}^{n}\tilde\xi_i=O_{P}(n^{-1/2})$.
 In addition, applying Lemma~\ref{le2} and
the triangle inequality, we obtain that
$$\bigl|n^{-1}\sum_{i=1}^{n}\hat\xi_i\bigr|
\leq \bigl|n^{-1}\sum_{i=1}^{n}(\hat\xi_i-\tilde\xi_i)\bigr|
 +\bigl|n^{-1}\sum_{i=1}^{n}\tilde\xi_i\bigr|=O_{P}(n^{-1/2}).$$
\hfill$\diamond$


\textbf{Proof of Lemma~\ref{le6}}. 
%
Since $n^{1/\gamma-1/2}=o(1)$,
there exists a sequence $\phi_n$ such that
 $\phi_n=o(n^{-1/\gamma})$ and
$n^{-1/2}=o(\phi_n)$.
Define
$\Lambda_n=\{\lambda:\, |\lambda|\leq \phi_n\}$.
Then, by  the result
$\max_{1\leq i\leq n}|\hat\xi_i|=O_{P}(n^{1/\gamma})$ in Lemma~\ref{le4}(ii),
we get $\max_{1\leq i\leq n,\lambda\in \Lambda_n}|\lambda\hat\xi_i|=o_{P}(1)$.
Let
$$\bar\lambda=\arg\min_{\lambda\in \Lambda_n}\sum_{i=1}^n\log(1+\lambda\hat\xi_i).$$
Then $\max_{1\leq i\leq n}|\bar{\lambda}\hat\xi_i|=o_{P}(1)$.
Using Taylor's expansion, with probability going to 1, we obtain that
\begin{eqnarray}
0\leq \sum_{i=1}^n\log(1+\bar\lambda\hat\xi_i)&=&\bar\lambda\sum_{i=1}^n\hat\xi_i
-\frac{\bar\lambda^2}{2}\sum_{i=1}^n\frac{\hat\xi_i^2}{(1+c_{i}^*\bar\lambda\hat\xi_i)^2}\nonumber\\
&\leq& |\bar\lambda|\Bigl|\sum_{i=1}^n\hat\xi_i \Bigr| - \bar\lambda^2c\sum_{i=1}^n\hat\xi_i^2  \label{eqr1}
\end{eqnarray}
for some constants $0\le c_{i}^*\le 1$ and $0<c\leq 1$.
By condition A3, $\text{Var}(\tilde\xi_1)>c_1$,
  and by Lemma~\ref{le4}(ii),
  $n^{-1}\sum_{i=1}^n|\tilde\xi_i|^2=E|\tilde\xi_1|^2+o_{P}(1)$.
  Then, applying Lemma~\ref{le2}(ii), we establishes that
\begin{equation}\label{eqr2}
n^{-1}\sum_{i=1}^n\hat\xi_i^2\geq c_1+o_{P}(1).
\end{equation}
By Lemma~\ref{le4}(iv), we have
$n^{-1}\sum_{i=1}^{n}\hat\xi_i=O_{P}(n^{-1/2})$,
which, combined with \eqref{eqr1}-\eqref{eqr2}, 
yields that
 $$\bar\lambda=O_{P}(n^{-1/2})=o_{P}(\phi_n).$$
  Thus,
   with probability tending to $1$,
  $\bar\lambda$ is in the interior
of $\Lambda_n$.
Since $\sum_{i=1}^n\log(1+\lambda\hat\xi_i)$ is concave,
$P(\hat\lambda=\bar\lambda)\to 1.$
Hence, $\hat\lambda=O_{P}(n^{-1/2}).$
\hfill$\diamond$


{\spacingset{1.2}

}

\end{document}